\documentclass[%
 reprint,
superscriptaddress,
 amsmath,amssymb,
 aps,
]{revtex4-2}

\usepackage{amssymb}
\usepackage{amsmath}
\usepackage{textcomp}
\usepackage{units}
\usepackage{color}
\usepackage{hyperref}
\usepackage{upgreek}
\usepackage{graphicx}
\usepackage{lineno} 
\usepackage{bm}%
\newcommand{\tens}[1]{\boldsymbol{\mathrm{#1}}}
\newcommand{\vect}[1]{\boldsymbol{#1}}
\newcommand{\dif}{\mathrm{d}}
\usepackage{amsthm}

\usepackage{graphicx}
\usepackage{dcolumn}
\usepackage{bm}

\begin{document}

\preprint{APS/123-QED}

\title{Large Amplitude Oscillatory Extension (LAOE) of dilute polymer solutions}

\author{Steffen M. Recktenwald}
 \affiliation{Micro/Bio/Nanofluidics Unit, Okinawa Institute of Science and Technology Graduate University, 1919-1 Tancha, Onna-son, Okinawa 904-0495, Japan}
 \email{s.recktenwald@oist.jp}

\author{Thomas P. John}
 \affiliation{Department of Chemical Engineering, The University of Manchester, Oxford Road M13 9PL, Manchester, United Kingdom}

\author{Amy Q. Shen}
 \affiliation{Micro/Bio/Nanofluidics Unit, Okinawa Institute of Science and Technology Graduate University, 1919-1 Tancha, Onna-son, Okinawa 904-0495, Japan}

\author{Robert J. Poole}
 \affiliation{School of Engineering, The University of Liverpool, Brownlow Street, L69 3GH Liverpool, United Kingdom}

\author{Cl\'audio P. Fonte}
 \affiliation{Department of Chemical Engineering, The University of Manchester, Oxford Road M13 9PL, Manchester, United Kingdom} 

 \author{Simon J. Haward}
 \affiliation{Micro/Bio/Nanofluidics Unit, Okinawa Institute of Science and Technology Graduate University, 1919-1 Tancha, Onna-son, Okinawa 904-0495, Japan}

\date{\today}

\begin{abstract}
This study presents an experimental framework for large amplitude oscillatory extension (LAOE) to investigate nonlinear material properties of complex fluids. Using a microfluidic optimized shape cross-slot extensional rheometer, we generate approximately homogeneous planar extensional flows driven by programmable syringe pumps operating in oscillatory or pulsatile sinusoidal modes. Micro-particle image velocimetry and simultaneous pressure drop measurements are employed to analyze the time-dependent flow field and elastic stress response. For Newtonian fluids, a linear relationship between the applied strain rate and pressure drop is observed across a wide range of oscillation amplitudes and frequencies. In contrast, dilute polymer solutions exhibit significant deviations, with excess pressure drops and divergence between average strain rates along extension and compression axes during the LAOE cycle. By spanning a broad range of Weissenberg and Deborah numbers, we identify unique Lissajous curves and critical conditions for the onset of nonlinearities under oscillatory extension. Numerical simulations, assuming homogeneous flow, underpin the experimental findings, validating the robustness of our microfluidic approach. This study demonstrates the utility of oscillatory extensional flows for probing the nonlinear rheological behavior of soft materials, offering quantitative insights into their extensional properties under nonlinear flow conditions.
\end{abstract}

\maketitle

\section{Introduction}
Oscillatory shear tests are widely used to characterize soft matter and complex fluids, including polymer melts and solutions, biological fluids, and food products. In particular, small amplitude oscillatory shear (SAOS) tests have become a canonical tool for probing the linear viscoelastic properties in the limit of small deformations~\cite{Ferry1947}. By applying small-magnitude sinusoidal strains and measuring the material's time-dependent stress response, SAOS provides insight into the relationship between a material’s microstructure and its rheological properties~\cite{Ferry1980, Bird1987, Tschoegl1989}. From these measurements, key material functions such as the linear viscoelastic moduli, \mbox{$G^\prime$} and \mbox{$G^{\prime\prime}$}, can be determined.

However, practical applications and most processing operations often involve large and rapid deformations, requiring quantifying nonlinear material properties to accurately predict material behavior. As a result, large amplitude oscillatory shear (LAOS) tests have emerged as a pivotal technique for studying and quantifying the nonlinear viscoelastic behavior of complex fluids~\cite{Fletcher1954, Hyun2002, Wilhelm2002, Hyun2011, Giacomin2011}. LAOS tests have been employed for microstructural assessment of polymeric systems~\cite{Kamkar2022}, in food rheology~\cite{Wang2022}, in electrospinning fluids~\cite{Jones2024}, and to predict complex behavior in human blood~\cite{Armstrong2022a}. At large strain amplitudes, the material response generally becomes nonlinear, resulting in a distorted stress signal that deviates from a sinusoidal wave, and the response can be interpreted as one containing higher-order harmonics. To characterize and understand this nonlinear stress response, several analysis methods have been developed, including power series expansions~\cite{Philippoff1966, Harris1967}, Fourier transform rheology~\cite{Wilhelm1998, Wilhelm1999}, stress decomposition~\cite{Cho2005} and decomposition into characteristic response functions~\cite{Klein2007, Klein2008}, Chebyshev polynomials~\cite{Ewoldt2008, Ewoldt2013}, the sequence of physical processes method~\cite{Rogers2012, Rogers2017, Park2018, Choi2019}, and recovery rheology~\cite{Lee2019a}.

While these techniques provide comprehensive frameworks for analyzing complex rheological behavior under large oscillatory shear deformations, the potential of oscillatory flows to reveal the extensional properties of complex fluids remains largely unexplored. In contrast to shear flow, where material elements separate linearly in time, extensional flow kinematics, where material elements separate exponentially in time, can induce significant microstructure deformation and are ubiquitous in various industrial applications, including filament spinning, inkjet printing, blow molding, extrusion, coating, and flow through porous media. Moreover, complex fluids can exhibit significantly different extensional properties despite displaying similar behavior under shear deformations~\cite{Oswald2019a, Abdelgawad2024}. Consequently, large amplitude oscillatory extension (LAOE) techniques have been developed to address this gap. Rasmussen et al.~\cite{Rasmussen2008} used a modified filament stretching rheometer (FSR) to impose large amplitude oscillatory elongation on a polystyrene melt at a temperature of \(120^{\circ} \textrm{C}\). This was achieved by extending the sample in an oscillatory manner, which was superimposed on a constant uniaxial elongational flow. During the experiment, the filament diameter was measured using a laser micrometer, and the elongational stress generated within the filament was calculated from the measured total force, revealing the transient stress response during elongation. Bejenariu~\textit{et al.}~\cite{Bejenariu2010} adapted the FSR technique to perform LAOE on soft poly(dimethylsiloxane) (PDMS) networks without any background flow. The authors observed a phase shift between stress and strain and reported deviations from the linear behavior of the measured elongational stress during the cycle. 

LAOE has also been used to study elastomers and rubbers, widely used in various industrial applications~\cite{Jiang2015, Dessi2017, Staropoli2022}. For instance, Dessi~\textit{et al.}~\cite{Dessi2017} investigated the response of filled elastomers subjected to a large-amplitude oscillatory uniaxial extension using a Sentmanat Extensional Rheometer (SER) fixture mounted on a strain-controlled rheometer. Their study revealed convex, banana-shaped Lissajous-Bowditch plots of the measured stress responses as a function of uniaxial strain. Additionally, the authors developed analytical expressions for the Fourier components of the stress response~\cite{Dessi2017}. More recently, LAOE has found applications in food science, where it has been used to correlate the nonlinear viscoelasticity of dough with product quality~\cite{Liu2023}.

Besides these advancements in modeling and measuring the oscillatory extension of relatively high-viscosity or elastic materials, such as polymer melts, rubbers, and polymeric networks in a uniaxial flow field, implementing LAOE for low-viscosity fluids poses greater experimental challenges. For fluids such as dilute polymer solutions, effects like gravitational sagging, necking, and capillary instabilities can lead to filament breakup in FSR devices~\cite{Anna2001c, McKinley2002a}. To overcome these limitations, stagnation point devices such as microfluidic cross-slots have been employed for extensional flow oscillatory rheometry (EFOR) of polymer solutions in quasi-steady state planar elongation~\cite{Odell2006, Haward2010b, Haward2011a}. In such cross-slots, the sample fluid is pumped into one pair of opposing channels and withdrawn from a second pair, resulting in a stagnation point at the center of the device. By combining EFOR measurements, where the periodic flow was realized by piezoelectric micro-pumps, with birefringence imaging, Odell~and~Carrington~\cite{Odell2006} could assess macromolecular strain by measuring optical birefringence during the period. Similar stagnation point flows have also been used to probe the shape dynamics of elastic capsules~\cite{Bryngelson2019a}, vesicles~\cite{Lin2021}, and the transient dynamics of single polymers in LAOE~\cite{Zhou2016, Zhou2016a}. Zhou~and~Schroeder~\cite{Zhou2016a} investigated the dynamics of single DNA molecules in large amplitude oscillatory extensional flows and revealed the molecules' varying compression, rotation, and extension during the LAOE cycle. The authors also reported a critical flow strength in terms of a Weissenberg number (Wi) for single polymers in LAOE at which a transition from linear to nonlinear stretching behavior occurs.

Despite investigations into the flow behavior of single polymer chains and highly viscous polymer melts and networks, little is known about the rheological response of viscoelastic polymer solutions to LAOE. In such dynamic flow fields with strong extensional components, polymer unraveling is expected to induce a strongly nonlinear stress response driven by the entropic elasticity that forces the polymer chains to relax. However, the stress response of viscoelastic polymer solutions to LAOE has not been experimentally measured, and such strong non-linearities associated with the finite extensibility of the polymers cannot be probed by conventional LAOS.

This study introduces an experimental approach to investigate the fluid response of dilute and viscoelastic polymer solutions under LAOE. We employ a microfluidic optimized shape cross-slot extensional rheometer (OSCER) device to generate a practically homogeneous planar extensional flow~\cite{Alves2008a, Haward2013a, Recktenwald2019}. Periodic flow through the OSCER geometry is driven by programmable syringe pumps operating either in an oscillatory mode or in a pulsatile mode with a constant background flow. The time-dependent flow field within the OSCER is analyzed using micro-particle image velocimetry, while the simultaneous pressure drop is measured to evaluate the fluid’s elastic stress response. First, we perform steady flow experiments to determine the critical conditions under which the flow becomes unstable and breaks symmetry. Next, we verify the experimental method by examining the time-dependent flow of a Newtonian fluid during LAOE, exploring a range of oscillation amplitudes and frequencies, and demonstrating the linearity between the applied strain rate and the measured total pressure drop. Subsequently, we turn to viscoelastic dilute polymer solutions, which exhibit pronounced deviations from Newtonian behavior at high Weissenberg numbers. These deviations are quantified through velocimetry and excess pressure drop measurements, revealing distinctive Lissajous curves (pressure drop versus strain rate) that vary with flow strength and probing frequency. Finally, we compare the oscillating and pulsatile sinusoidal LAOE modes, highlighting their differences, and we validate our experimental findings with numerical predictions. This study establishes a promising new methodology for characterizing complex fluids under controlled nonlinear transient flow conditions, providing insights into their behavior in dynamic and extensional environments.

\section{Material and Methods}
\subsection{Test fluids}
\begin{figure}
\centering
\includegraphics[width=0.5\textwidth]{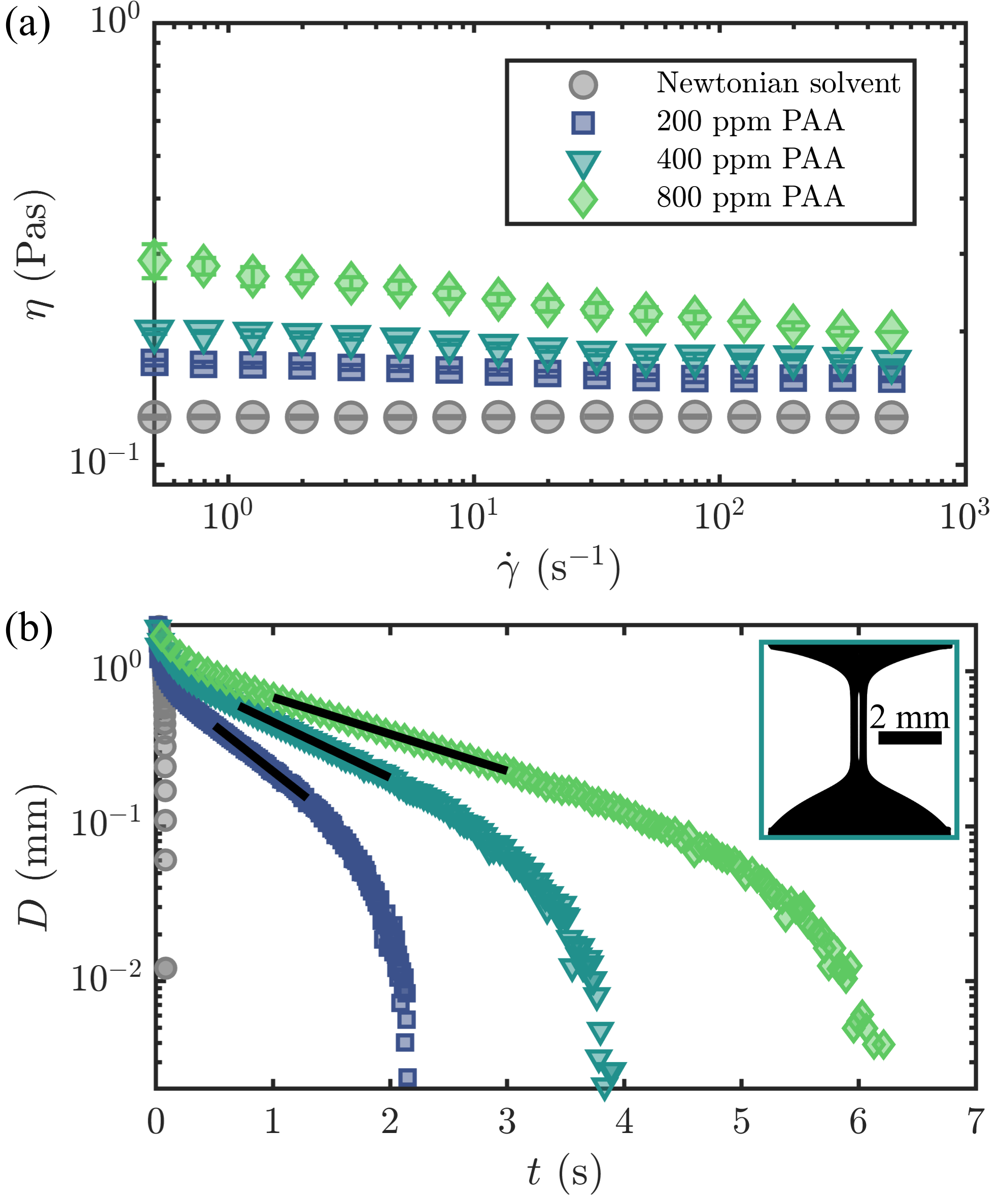}
\caption{Rheological responses of the test fluids in shear and uniaxial extension. (a) Shear viscosity as a function of the applied shear rate. (b) Representative curves of the thinning filament diameter during capillary thinning in a CaBER experiment. Black lines in (b) represent fits to the decaying diameter, used to extract the extensional relaxation times \mbox{$\lambda$} of the polymeric samples. The inset in (b) displays a representative snapshot of the cylindrical filament with a diameter of \mbox{$D=\unit[0.5]{mm}$}, captured during the elasto-capillary thinning regime for the 400~ppm PAA solution.}
\label{FIG_Fluids}
\end{figure}

In this work, we investigate the response of dilute polymer solutions to LAOE. As both a Newtonian reference fluid and a solvent for the viscoelastic polymeric solutions, we use a mixture consisting of \mbox{$\unit[89.6]{wt.\%}$} glycerol and \mbox{$\unit[10.4]{wt.\%}$} water. This mixture has a viscosity of \mbox{$\eta=\unit[128]{mPa\,s}$} (Fig.~\ref{FIG_Fluids}(a), Table\ref{tab:fluids}) and a density of \mbox{$\rho=\unit[1231]{kg/m^3}$}~\cite{Haward2023b}.

We use nonionic poly(acrylamide) (PAA, Sigma-Aldrich, Japan) with a molecular weight of \mbox{$M\approx\unit[5\times10^6]{g/mol}$} and prepare three polymer solutions at concentrations of \mbox{$c=200$}, \mbox{$400$}, and \mbox{$\unit[800]{\text{ppm}}$} (parts per million). The polymer powder is first dissolved in the aqueous component of the solvent and gently agitated on a roller mixer (Ika, Japan) for at least 24~h. Once the polymer is fully dissolved, glycerol is added, and the mixture is gently agitated for an additional 24~h. After preparation, the samples are stored at \mbox{$5^{\circ} \textrm{C}$} in the dark and discarded if unused within one month.

The overlap concentration \mbox{$c^*$} and the extensibility factor \mbox{$L$} for this polymer-solvent system have been previously estimated as \mbox{$\unit[c^*\approx 4400]{ppm}$} and \mbox{$L\approx 143$}, respectively~\cite{Haward2023b, Carlson2024}. Consequently, the polymer samples used in this study are in the dilute concentration regime (\mbox{$0.05 \lesssim c/c^* \lesssim 0.18$}).

\begin{table}[b]
\caption{\label{tab:fluids}%
Overview of the shear viscosity \mbox{$\eta$} and the time constant \mbox{$\lambda$} obtained from the rheological characterization of the test fluids. The values for \mbox{$\eta$} are derived from the high shear rate regime \mbox{$\dot{\gamma}=\unit[100-500]{s^{-1}}$}, while \mbox{$\lambda$} corresponds to the exponential filament decay measured during capillary thinning experiments.}
\centering
\begin{tabular}{lcc}
\hline
\textrm{Fluid}&
\textrm{\mbox{$\eta$} (\mbox{$\unit[]{mPa\,s}$})}&
\textrm{\mbox{$\lambda$} (\mbox{$\unit[]{s}$})}\\
\hline
Newtonian & 128 & - \\
200 ppm PAA & 157 & 0.25 \\
400 ppm PAA & 175 & 0.34 \\
800 ppm PAA & 209 & 0.58 \\
\hline
\end{tabular}
\end{table}

The steady shear rheology of the test fluids was measured using a stress-controlled DHR3 rotational rheometer (TA Instruments Inc., DE) equipped with a \mbox{$\unit[40]{mm}$} diameter \mbox{$1^{\circ}$} angle cone-and-plate geometry. The 200~ppm and 400~ppm PAA solutions exhibit a nearly constant viscosity, while the 800~ppm PAA solution shows slight shear-thinning behavior within the investigated shear rate range (Fig.~\ref{FIG_Fluids}(a)).

The fluid response under oscillatory shear was investigated using a strain-controlled rheometer (ARES G2, TA Instruments Inc., DE) equipped with a \mbox{$\unit[50]{mm}$} diameter \mbox{$1^{\circ}$} angle cone-and-plate geometry and a solvent trap. Strain amplitude sweeps were conducted for \mbox{$\gamma=\unit[0.5-500]{\%}$} at various frequencies, alongside a frequency sweep at \mbox{$\gamma=\unit[5]{\%}$} within the linear viscoelastic regime. Across all tested frequencies, \mbox{$G^{\prime\prime}\gg G^\prime$}, as shown in Fig.~S1(a) in the Supplementary Material. Furthermore, slightly elliptical Lissajous curves of shear stress versus shear rate reveal the dominant viscous behavior of the polymer samples under LAOS (Fig.~S1(b)).

The flow behavior of the fluid samples under uniaxial extension was characterized using a capillary breakup extensional rheometer (CaBER) device (Thermo-Haake, Germany). Circular plates with a diameter of \mbox{$\unit[6]{mm}$} and a strike time of \mbox{$\unit[200]{ms}$} were used to separate the plates from an initial gap of \mbox{$\unit[1]{mm}$} to a final gap of \mbox{$\unit[6]{mm}$}. The thinning filament diameter was monitored over time (Fig.~\ref{FIG_Fluids}(b)), and the time constant \mbox{$\lambda$} of the exponential filament decay (indicated by black lines in Fig.~\ref{FIG_Fluids}(b)) was extracted from the elasto-capillary thinning regime (see representative inset image in Fig.~\ref{FIG_Fluids}(b))~\cite{Anna2001a, Alves2008a}. Viscosity values at high shear rates (\mbox{$\dot{\gamma}=\unit[100-500]{s^{-1}}$}) and the extracted values of \mbox{$\lambda$} are summarized in Table~\ref{tab:fluids}.

\subsection{Microfluidic setup}
\begin{figure*}
\centering
\includegraphics[width=\textwidth]{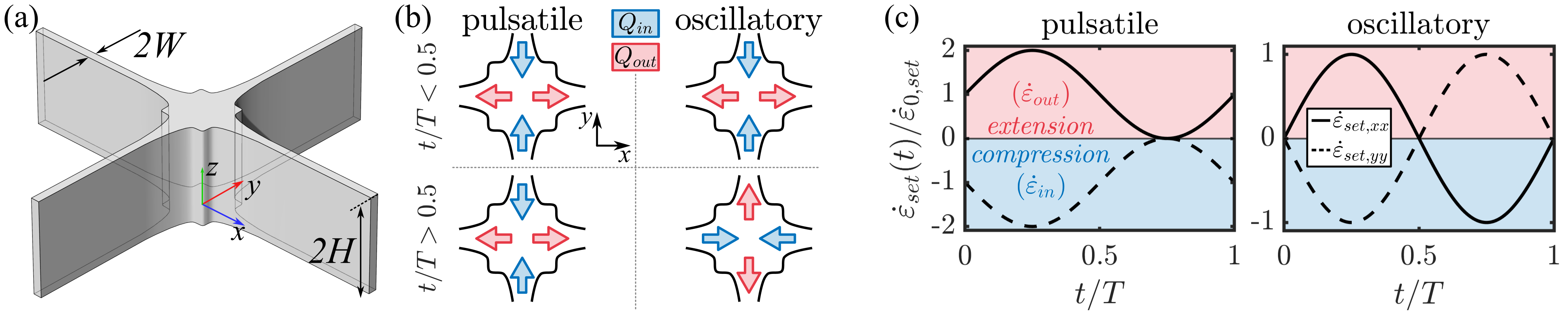}
\caption{Overview of the experimental setup and operating procedure. (a) Schematic illustration of the optimized shape cross-slot extensional rheometer (OSCER) geometry, featuring two pairs of opposing inlet and outlet channels aligned along the \mbox{$x$} and \mbox{$y$} axes, respectively. The inlet and outlet channels have a width of \mbox{$2W$}, and the geometry's height is \mbox{$2H$}. (b) Schematic depiction of the inlet \mbox{$Q_{in}$} and outlet \mbox{$Q_{flow}$} flow directions during (left) pulsatile and (right) oscillatory measurements at (top) \mbox{$t/T<0.5$} and (bottom) \mbox{$t/T>0.5$}. (c) Representation of the set extension rate profiles \mbox{$\dot{\varepsilon}_{set,xx}$} and \mbox{$\dot{\varepsilon}_{set,yy}$} in the \mbox{$x$} and \mbox{$y$} directions, normalized by the strain rate amplitude, for pulsatile (left) and oscillatory (right) modes. Shaded red and blue areas correspond to the fluid extension (\mbox{$\dot{\varepsilon}_{out}$}) and compression (\mbox{$\dot{\varepsilon}_{in}$}), respectively.}
\label{FIG_Setup}
\end{figure*}

We employ a microfluidic optimized shape cross-slot extensional rheometer (OSCER) geometry to generate planar extension~\cite{Alves2008a, Haward2013a, Recktenwald2019, Haward2023a}. The OSCER features a half-height of \mbox{$\unit[H=1]{mm}$} and a characteristic half-width of \mbox{$\unit[W=100]{\upmu m}$} for the inlet and outlet channels (Fig.~\ref{FIG_Setup}(a)). The geometry is fabricated from stainless steel using wire-electrical discharge machining, and sealed with glass viewing windows on the top and bottom. The high aspect ratio of the device (\mbox{$H/W=10$}) closely approximates a 2D flow, ensuring uniform flow across most of the channel height.

In the central region of the OSCER geometry (\mbox{$\lvert x \rvert,\lvert y \rvert \leq 15W$}), the flow approximates pure planar homogeneous elongation~\cite{Haward2012d, GalindoRosales2014, Haward2016a}. The OSCER device is mounted on an inverted microscope (Eclipse Ti, Nikon, NY) equipped with a \mbox{$4\times$} air objective (PlanFluor, Nikon, NY) with a numerical aperture of \mbox{$NA=0.13$}.

\subsection{Flow control}
\subsubsection{Steady flow conditions}
We perform measurements under both steady and time-dependent flow conditions. Below, we detail the experimental techniques employed for steady flow. The same methods are adapted for pulsatile and oscillatory flows, as described in Sec.~\ref{sec_timeMeasurements}.

For all measurements, the test fluids are driven through the microfluidic channel using low-pressure syringe pumps (Nemesys S, Cetoni GmbH, Germany) with a \mbox{$29:1$} gear ratio. The pumps are equipped with borosilicate glass syringes (Cetoni GmbH, Germany) of appropriate volumes (\mbox{$\unit[1-10]{mL}$}). The syringes are connected to the OSCER device via PTFE tubing (inner diameter \mbox{$\unit[1]{mm}$}, Darwin Microfluidics, France). Careful measures are taken to eliminate air bubbles in the channels, tubing, and syringes, as trapped bubbles can affect the system's response, particularly during time-dependent tests.

To achieve controlled flow, two pumps supply equal volumetric flow rates \mbox{$Q_{in}$} to the opposing inlet channels, while two additional pumps simultaneously withdraw fluid at equal and opposite rates \mbox{$Q_{out}=-Q_{in}$} from the outlet channels.

For steady flow conditions, the average flow velocity in each channel of the OSCER device is given by \mbox{$U=Q/(4WH)$}, where \mbox{$Q$} is the volumetric flow rate and \mbox{$W$} and \mbox{$H$} are the width and height of the device, respectively. The resulting set extension rate in the OSCER is \mbox{$\dot{\varepsilon}_{set}=0.1U/W$}~\cite{Haward2012d, Haward2013a, Haward2016a}. This set extension rate corresponds to the theoretical elongation rate experienced by a creeping Newtonian fluid along the stretching axis, which spans between the two outlet channels.

The average strain rate along the outflow direction is denoted as \mbox{$\dot{\varepsilon}_{out}$}, while the average strain rate along the inflow direction is labeled \mbox{$\dot{\varepsilon}_{in}$}. For a Newtonian fluid under 2D planar elongation, \mbox{$\dot{\varepsilon}_{out} = -\dot{\varepsilon}_{in}$}. Since the inlet and outlet directions align with the channel's \mbox{$x$} and \mbox{$y$} axes, we define \mbox{$\dot{\varepsilon}_{in} = \dot{\varepsilon}_{yy}$} and \mbox{$\dot{\varepsilon}_{out} = \dot{\varepsilon}_{xx}$} for measurements under steady flow conditions, where \mbox{$\dot{\varepsilon}_{yy} = -\dot{\varepsilon}_{xx}$}. Note that the strain rate in inlet direction \mbox{$\dot{\varepsilon}_{in}$} always corresponds to the strain rate along the compression axis, and the strain rate in outlet direction \mbox{$\dot{\varepsilon}_{out}$} corresponds to the strain rate along the extension axis in the OSCER geometry.

\subsubsection{Time-dependent flow conditions}\label{sec_timeMeasurements}
We investigate the fluid's response to oscillatory and pulsatile flow conditions. To generate these flow profiles, we program the syringe pumps to impose a sinusoidal extension rate profile

\begin{equation}
    \dot{\varepsilon}_{set}(t) = \dot{\varepsilon}_{off,set} + \dot{\varepsilon}_{0,set}\,\sin(2\pi\, t/T), 
    \label{eq_setsin} 
\end{equation}

\noindent where \mbox{$\dot{\varepsilon}_{off,set}$} represents the offset of the signal, corresponding to a constant background flow, \mbox{$\dot{\varepsilon}_{0,set}$} is the amplitude of the extension rate oscillation, and \mbox{$T$} is the oscillation period. The subscript \textit{set} refers to \textit{as programmed at the pump}. This set extension rate profile approximates the elongation experienced by the fluid along the stretching axis, which spans between the two outlet channels, assuming the set profiles are exactly reproduced. The set extension rate \mbox{$\dot{\varepsilon}_{set,yy}$} is applied as an input for two pumps at opposing inlet channels in the \mbox{$y$} direction, while two additional pumps impose a strain rate profile of \mbox{$\dot{\varepsilon}_{set,xx} = -\dot{\varepsilon}_{set,yy}$} at the two outlet channels.

For the pure oscillatory flow mode, the time-dependent extension rate oscillates around zero with no net flow, so we set \mbox{$\dot{\varepsilon}_{off,set}=0$}. While the offset strain rate can be set to any arbitrary value, we use \mbox{$\dot{\varepsilon}_{off,set}=\dot{\varepsilon}_{0,set}$} for all pulsatile measurements conducted in this study. Thus, the set input strain rate is described by the two parameters \mbox{$\dot{\varepsilon}_{0,set}$} and \mbox{$T$}, which cover a broad range of  \mbox{$\dot{\varepsilon}_{0,set}=\unit[0.5-50]{s^{-1}}$} and \mbox{$T=\unit[0.5-50]{s}$}, depending on the fluid under investigation.

Under unidirectional pulsatile flow, the fluid is injected with \mbox{$Q_{in}$} (or withdrawn with \mbox{$Q_{out}$}) along the \mbox{$y$} (or \mbox{$x$}) direction throughout the entire period (Fig.~\ref{FIG_Setup}(b), left), similar to the steady flow scenario. Therefore, the \mbox{$x$} axis is the extensional axis, and the \mbox{$y$} axis is the compressional axis during the entire cycle. Assuming the set profiles are exactly reproduced along the stretching axis, \mbox{$\dot{\varepsilon}_{in} = \dot{\varepsilon}_{set,yy} \leq 0$} and \mbox{$\dot{\varepsilon}_{out} = \dot{\varepsilon}_{set,xx} \geq 0$} during the entire cycle for pulsatile LAOE (Fig.~\ref{FIG_Setup}(c), left).

In the oscillatory mode, the pumps reverse direction at \mbox{$t/T = 0.5$}, causing the channels to switch between injection \mbox{$Q_{in}$} and withdrawal \mbox{$Q_{out}$} operation every half period (Fig.~\ref{FIG_Setup}(b), right). During the first half-cycle (\mbox{$0 < t/T < 0.5$}), the \mbox{$x$} axis remains the extensional axis (\mbox{$\dot{\varepsilon}_{set,xx} \geq 0$}) while the \mbox{$y$} axis is the compressional axis (\mbox{$\dot{\varepsilon}_{set,yy} \leq 0$}). During the second half-cycle (\mbox{$0.5 < t/T < 1$}), the roles reverse and the \mbox{$x$} axis becomes the compressional axis (\mbox{$\dot{\varepsilon}_{set,xx} \leq 0$}), and the \mbox{$y$} axis becomes the extensional axis (\mbox{$\dot{\varepsilon}_{set,yy} \geq 0$}) (Fig.~\ref{FIG_Setup}(c), right). This causes the extensional axis to alternate by \mbox{$90^\circ$} between the \mbox{$x$} and \mbox{$y$} axes every half period. The strain rate along the outflowing stretching axis (\mbox{$\dot{\varepsilon}_{out}$}) follows the modulus of the sinusoidal input, corresponding to \mbox{$\dot{\varepsilon}_{set,xx}$} for \mbox{$0 < t/T < 0.5$} and \mbox{$\dot{\varepsilon}_{set,yy}$} for \mbox{$0.5 < t/T < 1$}. Similarly, the inflowing compression strain rate (\mbox{$\dot{\varepsilon}_{in}$}) alternates between \mbox{$\dot{\varepsilon}_{set,yy}$} and \mbox{$\dot{\varepsilon}_{set,xx}$}, maintaining \mbox{$\dot{\varepsilon}_{in} = -\dot{\varepsilon}_{out}$}, as shown in Fig.~\ref{FIG_Setup}(c).

Precise synchronization of the pumps, pressure acquisition, and flow velocimetry is essential for time-dependent measurements. A global trigger signal, generated at the start of the sinusoidal flow modulation using a multi-function DAQ device (USB-6009, National Instruments, TX), ensures synchronization of all components. For pulsatile LAOE, a background flow at \mbox{$\dot{\varepsilon}_{0,set}$} is maintained for several seconds to stabilize the flow before the superimposed oscillatory modulation begins.

\subsection{Microparticle image velocimetry}
We measure the flow field in the OSCER using micro-particle image velocimetry ($\upmu$-PIV, TSI Inc., MN)~\cite{Wereley2010, Wereley2019}. The test fluids are seeded with \mbox{$\unit[0.02]{wt.\%}$} of \mbox{$\unit[2]{\upmu m}$} red fluorescent tracer particles (Fluor-Max, Thermo Scientific, Germany), with excitation and emission wavelengths of \mbox{$\unit[542]{nm}$} and \mbox{$\unit[612]{nm}$}, respectively. A volumetric illumination technique is used, and the tracer particles are excited with a dual-pulsed Nd:YLF laser (\mbox{$\unit[527]{nm}$}). Flow is recorded in the \mbox{$xy$} midplane of the OSCER geometry using a high-speed camera (Phantom MIRO, Vision Research, Canada). The depth over which the tracer particles contribute to the velocity field measurement is \mbox{$\unit[\delta_z \approx161]{\upmu m}$}~\cite{Meinhart2000} (\mbox{$\delta_z\approx0.16 H$}).

The high-speed camera operates in frame-straddling mode, synchronized with the laser. The frame rate and the separation of laser pulses \mbox{$\Delta t$} are adjusted based on the applied flow rate and used fluid to achieve an average particle displacement of approximately 4 pixels between image pairs. 

For steady flow measurements, between 50 and 200 image pairs are recorded. Image pre-processing is performed to subtract the image background, and then PIV analysis is conducted (TSI Insight 4G, TSI Inc., MN). This analysis includes ensemble averaging over the image sequence, employing a recursive Nyquist criterion with an interrogation window size of \mbox{$16\times16$} pixels.

During time-dependent PIV measurements, the laser pulse duration is constant to achieve a tracer particle displacement of approximately 4~pixels at the maximum strain rate (\mbox{$\dot{\varepsilon}_{off,set}+\dot{\varepsilon}_{0,set}$}) during the cycle. The camera frame rate is set to capture at least 100 image pairs per cycle, with coverage ranging from 100 image pairs for \mbox{$T = \unit[0.5]{s}$} to 1300 for \mbox{$T = \unit[50]{s}$}. At least 4 full oscillation cycles are recorded per measurement for longer periods (\mbox{$T = \unit[50]{s}$}), increasing to 12 cycles for shorter periods (\mbox{$T = \unit[0.5]{s}$}).

The velocity components \mbox{$u$} and \mbox{$v$} in the \mbox{$x$} and \mbox{$y$} directions, respectively, are obtained from the velocity field \mbox{$\mathbf{u}$}. Subsequent data analysis uses a custom \textsc{MATLAB} (R2024a, The MathWorks, MA) algorithm.

\subsection{Pressure drop measurements}\label{sec_pressuremeasure}
We measure the pressure drop across one inlet and one outlet channel of the OSCER device using various wet-wet differential pressure transducers (\mbox{$\unit[6.9-35]{kPa}$}, Omega Engineering Inc., Germany). Pressure taps are installed on the tubing via T-junctions between the syringes and the OSCER device for both the inlet and outlet channels. For each sample and set extension rate, two independent pressure drop measurements are performed, as described in previous studies~\cite{Haward2011a, Haward2023b}.

The first measurement captures the total pressure drop \mbox{$\Delta P_{tot}$} with flow imposed in all four channels. The second measurement is conducted with flow restricted to only the two channels connected to the pressure transducers, while the other two channels are disabled. This configuration allows us to measure the pressure drop associated with the flow of fluid around a corner in the cross-slot, \mbox{$\Delta P_{sh}$}. The excess pressure drop, \mbox{$\Delta P_{ex} = \Delta P_{tot} - \Delta P_{sh}$}, is then calculated and used to quantify the additional stresses resulting from the elongational kinematics of the flow~\cite{Haward2011a, Haward2012a, Haward2023b}.

For pressure drop measurements (\mbox{$\Delta P_{tot}$} and \mbox{$\Delta P_{sh}$}) during pulsatile and oscillatory flow, the pressure signal is recorded over at least 10 cycles, ranging from 10 cycles for \mbox{$T = \unit[50]{s}$} to 120 cycles for \mbox{$T = \unit[0.5]{s}$}. Each pressure and PIV measurement is repeated three times for accuracy. Additionally, time-dependent pressure and velocity data are phase-averaged to compute the excess pressure and derive the corresponding Lissajous curves.

\subsection{System characteristics}\label{sec_methodsystem}
Applying a time-dependent flow rate in microfluidic systems can lead to significant deviations between the set input signal \mbox{$\dot{\varepsilon}_{set}(t)$} and the actual inlet signal \mbox{$\dot{\varepsilon}_{in}(t)$} within the microfluidic chip, particularly at high frequencies (see Fig.~S2(a,b) in the Supplementary Material). To understand these deviations, we investigate the system's frequency response, which includes the pumps, syringes, tubing, and the connected microfluidic chip, at various driving amplitudes and frequencies. 

\begin{figure}
\centering
\includegraphics[width=0.5\textwidth]{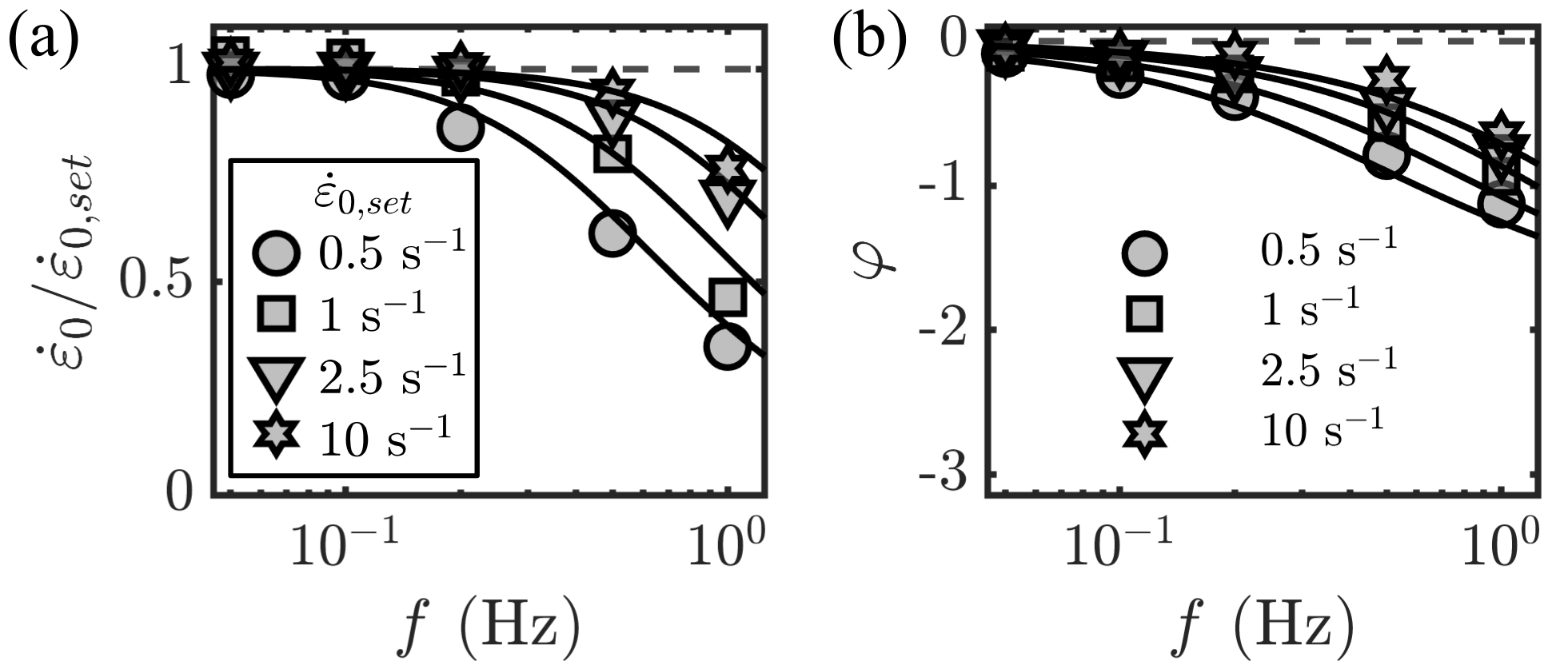}
\caption{System response under pulsatile LAOE. (a) Ratio of the measured inlet extension rate amplitude to the set amplitude, \mbox{$\dot{\varepsilon}_{0}/\dot{\varepsilon}_{0,set}$}, and (b) the phase shift, \mbox{$\varphi$}, between the inlet extension profile and the set profile, as functions of the pulsation frequency. Solid black lines represent fit curves, as detailed in the Supplementary Material.}
\label{FIG_System}
\end{figure}

Figure~\ref{FIG_System} shows representative Bode plots of (a) the strain rate amplitude ratio \mbox{$\dot{\varepsilon}_{0}/\dot{\varepsilon}_{0,set}$} and (b) the phase shift \mbox{$\varphi$} as a function of frequency, for various \mbox{$\dot{\varepsilon}_{0,set}$} under pulsatile LAOE. Figure~S2(c) in the Supplementary Material shows representative Bode plots under oscillatory LAOE. At low frequencies, the set amplitude is closely followed during the cycle with \mbox{$\dot{\varepsilon}_{0}/\dot{\varepsilon}_{0,set}=1$} and \mbox{$\varphi=0$}, as indicated by the horizontal dashed lines in Fig.~\ref{FIG_System}. However, as the frequency increases, the amplitude ratio decreases continuously, and a phase shift emerges between the set input signal and the inlet strain rate profile in the OSCER. Moreover, this effect becomes more pronounced as the set amplitude \mbox{$\dot{\varepsilon}_{0,set}$} decreases at a fixed pulsation frequency (see Fig.~S2(a,b)). The dependence of the strain rate amplitude ratio \mbox{$\dot{\varepsilon}_{0}/\dot{\varepsilon}_{0,set}$} and the phase shift \mbox{$\varphi$} can be described by linear response theory (see black lines in Fig.~\ref{FIG_System}), providing a characterization of the microfluidic system, similar to previous studies~\cite{VanDerBurgt2014, Li2018, Recktenwald2021b}. Details about modeling the linear system response are provided in the Supplementary Material.

We find that the microfluidic system used in this study is overdamped, resulting in significant differences between the set input signal \mbox{$\dot{\varepsilon}_{set}(t)$} and the inlet strain rate in the OSCER \mbox{$\dot{\varepsilon}_{in}(t)$}. Based on the transfer function of the model system and the frequency-response data (see Fig.~\ref{FIG_System}), an adapted set extension rate profile can be calculated, which significantly reduces deviations between the desired and actual signals. This approach has been demonstrated for pressure-driven microfluidic systems~\cite{Recktenwald2021b}. In this study, we focus on the transient pressure response to periodic driving without implementing an adapted input signal in the syringe pumps. Instead, we address amplitude attenuation and phase shift by directly measuring the actual flow profile in the cross-slot using PIV, rather than relying on \mbox{$\dot{\varepsilon}_{set}(t)$}. This approach also enables us to assess whether flow modifications affect the local flow field within the OSCER, which cannot be determined \textit{a priori}. Notably, the most accurate approximations to a sinusoidal modulation of the imposed strain rate profile, with minimal deviations from the set extension rate and phase shift in the OSCER, are observed at low frequencies or high strain rate amplitudes.

\subsection{Data normalization and dimensionless groups}
Due to the aspects observed in Sec.~\ref{sec_methodsystem}, we treat the temporal inlet extension rate \mbox{$\dot{\varepsilon}_{in}(t)$} measured by PIV as imposed driving signal, which is fitted to a sinusoidal function, \mbox{$\dot{\varepsilon}_{in}^{fit}(t)=\dot{\varepsilon}_{off}+\dot{\varepsilon}_{0} \sin(2\pi\, t/T)$}. From this fit, the offset \mbox{$\dot{\varepsilon}_{off}$} and amplitude \mbox{$\dot{\varepsilon}_{0}$} of the measured signal are extracted. Additionally, the phase shift \mbox{$\varphi$} between the inlet strain rate profile \mbox{$\dot{\varepsilon}_{in}(t)$} and the set signal \mbox{$\dot{\varepsilon}_{set}(t)$} is determined. Based on the offset \mbox{$\dot{\varepsilon}_{off}$} and amplitude \mbox{$\dot{\varepsilon}_{0}$}, the extension rates are normalized by the maximum of the inlet strain rate profile \mbox{$\dot{\varepsilon}_{max}=\dot{\varepsilon}_{0}+\lvert \dot{\varepsilon}_{off} \rvert$}, as \mbox{$\dot{\varepsilon}_{out}^\prime= \lvert \dot{\varepsilon}_{out}(t)/\dot{\varepsilon}_{max} \rvert$} and \mbox{$\dot{\varepsilon}_{in}^\prime= \lvert \dot{\varepsilon}_{in}(t)/\dot{\varepsilon}_{max} \rvert$}. Similarly, the pressure profile is fitted to a sinusoid, and the pressure signal is normalized by the maximum pressure during the cycle as \mbox{$\Delta P_{tot}^\prime= \lvert \Delta P_{tot}(t)/\Delta P_{tot,max} \rvert$}. This normalization is also applied to oscillatory measurements (\mbox{$\dot{\varepsilon}_{off}\equiv0$}) and is used to plot Lissajous figures of inlet versus outlet strain rates and pressure drop versus extension rates. 

The Reynolds number (\mbox{$\text{Re}$}), which describes the ratio of inertial to viscous forces during flow, is calculated as \mbox{$\text{Re} = \rho U D_h / \eta$}, where \mbox{$\rho$} is the fluid density, \mbox{$D_h = 2WH/(W+H)$} is the hydraulic diameter of the rectangular inlet and outlet channels, and \mbox{$\eta$} is the dynamic viscosity. The maximum Reynolds number reached in this study is \mbox{$\text{Re} \approx 0.07$}, indicating that inertial effects are negligible in all experiments.

The Weissenberg number quantifies the relative importance of elastic versus viscous stresses during flow. For steady flow conditions, the Weissenberg number is calculated as \mbox{$\text{Wi} = \lambda \dot{\varepsilon}_{set}$}, where \mbox{$\lambda$} is the polymeric relaxation time obtained from CaBER measurements. To characterize elastic effects under time-dependent measurements, we define the maximum Weissenberg number \mbox{$\text{Wi}_{max} = \lambda \dot{\varepsilon}_{max}$}.

For time-dependent measurements, we define the Deborah number De, defined as the ratio of the fluid's relaxation time \mbox{$\lambda$} to the oscillation period \mbox{$T$}~\cite{Poole2012}, \mbox{$\text{De} = \lambda / T$} for pulsatile LAOE under unidirectional flow conditions. Under oscillatory LAOE, the fluid is effectively extended during two periods of duration \mbox{$T/2$} each, once along the \mbox{$x$} direction (for \mbox{$0<t/T<0.5$}) and once along the \mbox{$y$} direction (for \mbox{$0.5<t/T<1$}) within one full imposed cycle (see Fig.~\ref{FIG_Setup}(c)). As a result, the effective period duration for oscillatory LAOE is half the period length \mbox{$T$} of the imposed sinusoidal signal \mbox{$\dot{\varepsilon}_{set}(t)$}, and the Deborah number for oscillatory LAOE measurements is calculated as \mbox{$\text{De}=2\lambda/T$}. 

\subsection{Simulations}\label{sec_methodsim}
We compare to the experimental data the responses of two commonly used non-linear viscoelastic constitutive models under oscillatory spatially-homogeneous extensional flow, namely the FENE-P \cite{Bird1987} (Finitely Extensible Nonlinear Elastic with Peterlin closure approximation) and Giesekus \cite{Giesekus1982} models. These are described by a generalized model given in dimensionless form as

\begin{equation}
\text{De} \frac{\dif \tens{A}}{\dif t} -\text{Wi}_\text{max} \left[ \tens{A}\vect{\cdot}\vect{\nabla}\vect{u} +{\vect{\nabla}\vect{u}}^{\mathrm{T}}\vect{\cdot}\tens{A} \right] = \left[ \tens{I}+\alpha(\tens{A}-\tens{I}) \right]\vect{\cdot}(f\tens{A}-a\tens{I}), \label{eq:constitutivemodel}
\end{equation}

\noindent where $\tens{A}$ is the conformation tensor, $L^2$ is the FENE extensibility limit, and $\alpha$ is the Giesekus mobility parameter. $f$ and $a$ are defined by $f \equiv L^2/(L^2 - \mathrm{tr}(\tens{A}))$ and $a \equiv L^2/(L^2-3)$, respectively. Note that the FENE-P model corresponds to the case where $\alpha = 0$ and $f > 1$ (i.e. $L^2 > 3$) and the Giesekus model corresponds to the case where $0 < \alpha \leq 1$ and $f = a = 1$.  We use an extensibility factor of $L=143$ \cite{Haward2023b, Carlson2024}, and we set $\alpha = 1/L^2$. Note that setting $\alpha = 1/L^2$ ensures that the extensional viscosity asymptotes towards the same high strain-rate limit for both the FENE-P and Giesekus models in a steady and homogeneous planar extensional flow~\cite{Poole2024}. The dimensionless extra-stress $\tens{\tau}$ tensor is recovered by

\begin{equation}
\tens{\tau} = \frac{(1-\beta)}{\text{Wi}_\text{max}}\left(f \tens{A} - a \tens{I} \right) + \beta(\vect{\nabla}\vect{u}+{\vect{\nabla}\vect{u}}^{\mathrm{T}}), \label{eq:stressconversion}
\end{equation}

\noindent where $\beta$ is the viscosity ratio (solvent to zero-shear). 

We consider a 2D homogeneous extensional flow field under a pulsatile driving mode, given in dimensionless form by 

\begin{eqnarray}
\vect{\nabla}\vect{u} & = & \dot{\varepsilon}(t) \begin{bmatrix} \displaystyle{\frac{1}{2}} & 0 & 0 \\ 0 & -\displaystyle{\frac{1}{2}} & 0 \\ 0 & 0 & 0 
\end{bmatrix}_{xyz} \\  & = & \nonumber \displaystyle{\frac{1}{2}}(1+\sin(2\pi t)) \begin{bmatrix} \displaystyle{\frac{1}{2}} & 0 & 0 \\ 0 & -\displaystyle{\frac{1}{2}} & 0 \\ 0 & 0 & 0 
\end{bmatrix}_{xyz}.
\label{eq:flowfield}
\end{eqnarray}

\noindent We substitute the flow field (Equation \eqref{eq:flowfield}) into the constitutive model (Equations \eqref{eq:constitutivemodel} and \eqref{eq:stressconversion}) and solve the resulting system of ODEs with \textsc{MATLAB}'s \textit{ode15s} solver from an equilibrium initial condition of $[A_{xx}, A_{yy}, A_{zz}]_{t=0} = [1, 1,1]$, i.e. $\tens{A} = \tens{I}$. The computed first normal stress difference ($N_1 \equiv \tau_{xx}-\tau_{yy}$) is normalized in the same way as the experimental results in order to make the comparison between experimental and numerical results. The full system of ODEs for each model is given in \ref{fullodes}.


\section{Results and discussion}
First, we analyze the flow behavior in the OSCER device under steady flow conditions in Sec~.\ref{sec_steady}, focusing on the limiting Weissenberg number values beyond which flow instabilities arise, as reported in prior studies~\cite{Arratia2006, Poole2007, Haward2013c, Haward2016a}. Next, we investigate pulsatile and oscillatory flow conditions in Secs.\ref{sec_pulse} and \ref{sec_osc}, respectively. For each time-dependent flow mode, we first present the behavior of the Newtonian reference fluid, followed by the results for non-Newtonian fluids. A direct comparison of the two flow modes is provided in Sec.~\ref{sec_compare}, while Sec.~\ref{sec_compareSim} contrasts experimental findings with numerical simulations.

\subsection{Steady flow}\label{sec_steady}
\subsubsection{Flow field characterization}
We begin by summarizing the flow characteristics of the Newtonian reference fluid in the OSCER under steady flow conditions. Within the investigated strain rate range (\mbox{$\dot{\varepsilon}_{set}=\unit[0.5-50]{s^{-1}}$}, \mbox{$\text{Re}<0.07$}), the flow field remains symmetric about the central stagnation point and the principal flow axes, as exemplified in Fig.~\ref{FIG_SteadyFlow}(a) and Fig.~S3(a-c) in the Supplementary Material. The velocity components \mbox{$v$} (inlet direction) and \mbox{$u$} (outlet direction) are extracted along their respective axes, indicated by the dashed lines in Fig.~\ref{FIG_SteadyFlow}(a). Over the range of flow rates studied, \mbox{$u$} increases, and \mbox{$v$} decreases along their respective axes (Fig.~S3(d) in the Supplementary Material).

For steady-state measurements, we calculate the average inlet (\mbox{$\dot{\varepsilon}_{in}=\partial v/\partial y$}) and average outlet (\mbox{$\dot{\varepsilon}_{out}=\partial u/\partial x$}) extension rates by averaging the velocity gradient over the spatial domain \mbox{$\lvert y/W \rvert, \lvert x/W \rvert \leq 6$}. Representative results for the Newtonian reference fluid at \mbox{$\dot{\varepsilon}_{set}=\unit[50]{s^{-1}}$} are shown in Fig.~\ref{FIG_SteadyFlow}(b). These findings are consistent with previous investigations of steady planar extension of Newtonian fluids~\cite{GalindoRosales2014, Haward2016a, Haward2023b}. The inlet and outlet extension rates correspond to the compression and extension axes of the investigated sample, respectively. 

\begin{figure*}
\centering
\includegraphics[width=\textwidth]{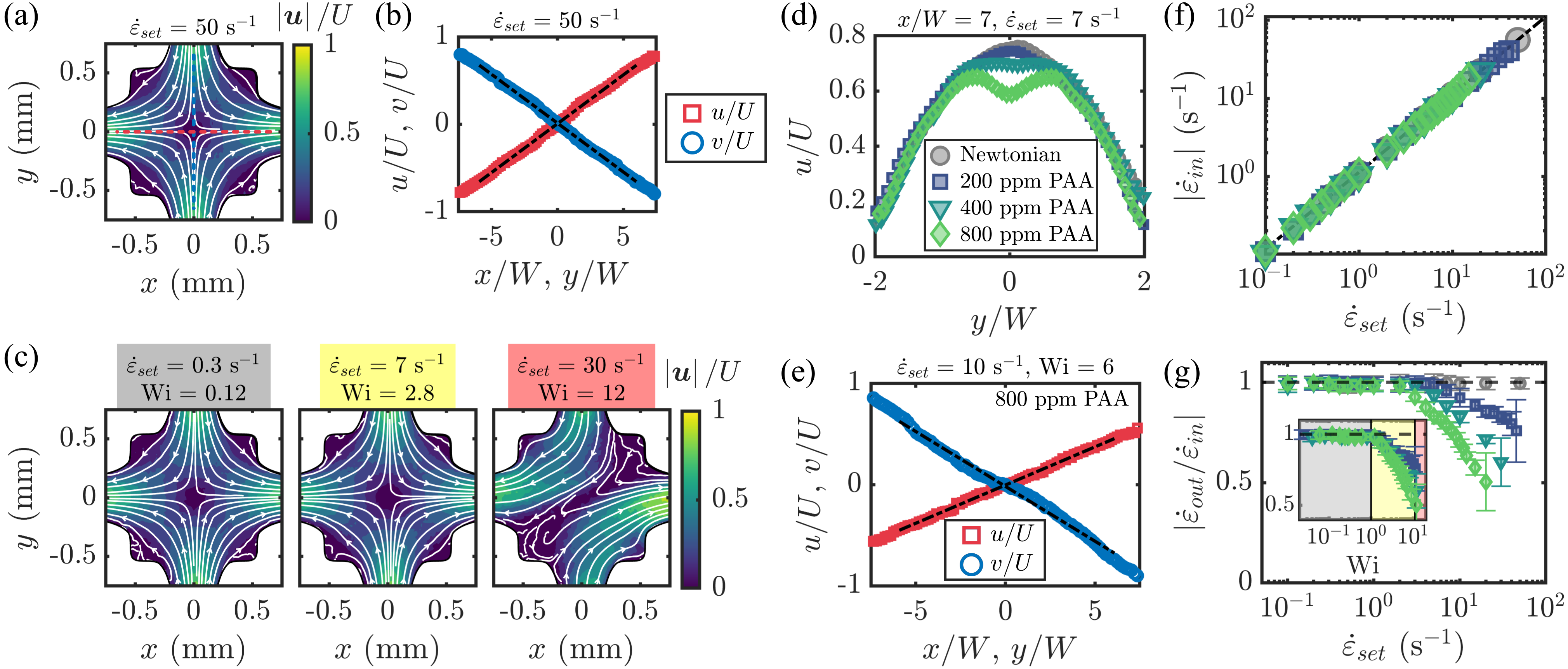}
\caption{Planar Newtonian and non-Newtonian flow under steady flow conditions. (a) Normalized velocity field with superimposed streamlines for the Newtonian fluid at \mbox{$\dot{\varepsilon}_{set}=\unit[50]{s^{-1}}$}. (b) Normalized velocity components along the $x$ and $y$ directions for the data in (a). Black dashed lines represent linear fits over \mbox{$\lvert x/W \rvert, \lvert y/W \rvert \leq 6$}, used to calculate extension rates. (c) Normalized velocity fields with streamlines at different \mbox{$\dot{\varepsilon}_{set}$} for the 400~ppm PAA sample. (d) Streamwise velocity profiles normalized by the centerline velocity for all tested solutions, measured across an outlet \mbox{$\unit[7]{mm}$} downstream of the stagnation point at \mbox{$\dot{\varepsilon}_{set}=\unit[7]{s^{-1}}$}. (e) Normalized velocity components for the 800~ppm PAA sample at \mbox{$\dot{\varepsilon}_{set}=\unit[10]{s^{-1}}$} (\mbox{$\text{Wi}=6$}). (f) Magnitude of measured inlet extension rate \mbox{$\lvert\dot{\varepsilon}_{in}\rvert$} vs. the set elongation rate for \mbox{$\text{Wi}<\text{Wi}{c}$}. (g) The magnitude of the ratio \mbox{$\lvert\dot{\varepsilon}_{out}/\dot{\varepsilon}_{in}\rvert$} as a function of the set extension rate, with inset showing this ratio as a function of \mbox{$\text{Wi}$}. Inset shading (grey, yellow, red) corresponds to symmetric flow, onset of asymmetry, and strongly asymmetric flow states, respectively, as seen in (c).}
\label{FIG_SteadyFlow}
\end{figure*}

For the investigated polymer solutions, the flow remains symmetric with a central stagnation point at \mbox{$\text{Wi}<1$}, resembling the Newtonian case, as shown for the 400~ppm PAA solution at \mbox{$\text{Wi}=0.12$} in Fig.~\ref{FIG_SteadyFlow}(c). With increasing Wi, the flow field initially retains its symmetry (\textit{e.g.}, at \mbox{$\text{Wi}=2.8$}). However, above a critical Weissenberg number, \mbox{$\text{Wi}_{inst}\approx11$}, we observe the onset of an elastically driven instability (\mbox{$\text{Re}\ll1$}). This results in complete symmetry breaking of the flow field in the OSCER, as seen at \mbox{$\text{Wi}=12$} in Fig.~\ref{FIG_SteadyFlow}(c). Similar flow asymmetries have been reported in prior studies on the steady planar extension of polymeric fluids and have been attributed to a purely elastic phenomenon driven by elastic tensile stresses~\cite{Gardner1982, Pakdel1996, Arratia2006, Poole2007, Rocha2009, Haward2012, Haward2013e, Haward2016a, Recktenwald2019}. In this study, we focus on probing the LAOE response of the polymeric test fluids within the symmetric flow regime at \mbox{$\text{Wi}<\text{Wi}_{inst}$}.

Under steady flow in the range \mbox{$1<\text{Wi}<\text{Wi}_{inst}$}, we observe significant modifications of the flow along the stretching axis, while the overall flow field remains symmetric. This effect is evident in the velocity profiles of the polymeric fluids across the channel outlet, measured \mbox{$\unit[0.7]{mm}$} downstream of the stagnation point, as shown for all tested fluids at \mbox{$\dot{\varepsilon}_{set}=\unit[7]{s^{-1}}$} in Fig.~\ref{FIG_SteadyFlow}(d). At this extension rate, the velocity profile of the 200~ppm PAA solution matches that of the Newtonian reference fluid, displaying a parabolic shape. In contrast, the 400~ppm solution exhibits a noticeable flattening of the velocity profile around \mbox{$y=0$}. For the 800~ppm solution, this modification becomes more pronounced, developing a distinct local minimum in the velocity profile at \mbox{$y=0$}.

This decrease in velocity near the stretching axis was previously reported in the literature~\cite{Lyazid1980, Gardner1982, Dunlap1987, Harlen1990, Haward2010, Haward2012d} and is attributed to localized polymer stretching along the outlet centerline. This stretching results in a higher extensional viscosity in the polymer samples compared to the surrounding fluid outside the stretching region. For polymer solutions exhibiting measurable flow-induced birefringence, this localized stretching also manifests as a distinct birefringent strand along the stretching axis~\cite{Keller1985, Becherer2008, Becherer2009, Sharma2015, Haward2012d, Haward2013e, Recktenwald2019}.

Similar to the Newtonian fluids, we calculate the extension rates \mbox{$\dot{\varepsilon}_{in}$} and \mbox{$\dot{\varepsilon}_{out}$} for the polymeric fluids by averaging the velocity gradients along the inlet and outlet directions, as representatively shown for \mbox{$\dot{\varepsilon}_{set}=\unit[10]{s^{-1}}$}, \mbox{$\text{Wi}=6$} in Fig.~\ref{FIG_SteadyFlow}(e). For \mbox{$\text{Wi}>1$}, we observe a kink in the normalized velocity profile \mbox{$v/U$} along the compression direction, located near the stagnation point at \mbox{$-2 \lesssim y/W \lesssim 2$}. In contrast, normalized velocity profile \mbox{$v/U$} along the extension direction shows a linear increase, similar to the Newtonian case or when \mbox{$\text{Wi}<1$} without the appearance of any kinks. Despite this modification in the velocity profile \mbox{$v/U$} along the inlet direction, when plotting the average extension rate \mbox{$\dot{\varepsilon}_{in} = \partial v/\partial y$} as a function of the set extension rate, the polymer solutions exhibit behavior that follows the Newtonian trend for all tested flow rates when \mbox{$\text{Wi} < \text{Wi}_{inst}$} (Fig.~\ref{FIG_SteadyFlow}(f)), in agreement with previous studies~\cite{Haward2023b}. 

Moreover, we observe a decrease in the average outlet extension rate relative to the average inlet extension rate for the polymeric fluids as the flow rate increases. To quantify this effect, we calculate the ratio \mbox{$\lvert\dot{\varepsilon}_{out}/\dot{\varepsilon}_{in}\rvert$} as a function of \mbox{$\dot{\varepsilon}_{set}$}, as shown for all tested fluids in Fig.~\ref{FIG_SteadyFlow}(g). This ratio remains constant and equal to 1 for the Newtonian reference fluid and for the polymer solutions at \mbox{$\text{Wi}<1$} (see inset in Fig.~\ref{FIG_SteadyFlow}(g)), indicating that \mbox{$\dot{\varepsilon}_{out} = \lvert\dot{\varepsilon}_{in}\rvert$}. However, at \mbox{$\text{Wi}_{c} \approx 1$}, the ratio begins to decrease, meaning \mbox{$\dot{\varepsilon}_{out} < \lvert\dot{\varepsilon}_{in}\rvert$} at their respective axes. This phenomenon is linked to the development of the local minimum in the velocity profile \mbox{$u(y)$} at the fluid outlet (see Fig.~\ref{FIG_SteadyFlow}(d)), where the extension rate is calculated along the \mbox{$x$}-direction at \mbox{$y=0$}. Consequently, the observed modification of the flow for the non-Newtonian fluids leads to a difference between the inlet and outlet extension rates for \mbox{$\text{Wi} \geq \text{Wi}_{c}\approx1$}.

Note that by virtue of mass conservation for incompressible materials (\textit{i.e.}, \mbox{$\partial U/\partial x +\partial V/\partial y=0$}), the strain rates in both axes are locally equal at the stagnation point.
 
\subsubsection{Pressure drop measurements}
We study the pressure drop of the test fluids in the OSCER under steady flow conditions, as summarized in Fig.~S4 of the Supplementary Material. In brief, we measure the total pressure drop \mbox{$\Delta P_{tot}$} and the shear-dominated pressure drop \mbox{$\Delta P_{sh}$} by progressively increasing \mbox{$\dot{\varepsilon}_{set}$} (Fig.~S4(a)). The steady-state plateau pressure drop is then measured at each strain rate step (Fig.~S4(b)). For polymer solutions at low extension rates, \mbox{$\Delta P_{tot} \approx \Delta P_{sh}$}, similar to the Newtonian reference fluid. At higher \mbox{$\dot{\varepsilon}_{set}$}, \mbox{$\Delta P_{tot}$} increases above \mbox{$\Delta P_{sh}$}. We calculate the excess pressure drop \mbox{$\Delta P_{ex} = \Delta P_{tot} - \Delta P_{sh}$} as a function of the extension rate (Fig.~S4(c)). We observe that the polymer solutions show a linear behavior of \mbox{$\Delta P_{ex}(\dot{\varepsilon}_{set})$} at low extension rates, similar to that of Newtonian fluids. However, the excess pressure drop at higher extension rates exceeds the linear trend at \mbox{$\text{Wi}>1$} and approaches plateau values, in agreement with previous studies~\cite{Haward2023b}.

In summary, we characterized the test fluids' flow field and pressure response under steady planar extensional flow. As shown in Fig.~\ref{FIG_SteadyFlow}, the flow of polymeric solutions is Newtonian-like at \mbox{$\text{Wi}<1$} with \mbox{$\dot{\varepsilon}_{out} = -\dot{\varepsilon}_{in}$}. At \mbox{$\text{Wi}_{c}=1$}, the response deviates from the Newtonian behavior, evidenced by a modification of the flow along the stretching axis and \mbox{$\dot{\varepsilon}_{out} < -\dot{\varepsilon}_{in}$}. Furthermore, we identified the critical \mbox{$\text{Wi}_{inst}\approx11$} at which the flow becomes unstable and asymmetric. Consequently, the time-dependent LAOE measurements are conducted at \mbox{$\text{Wi}_{max} < \text{Wi}_{inst}$}, and we expect a significant excess pressure drop to emerge above \mbox{$\text{Wi}_{c}=1$} in the polymeric samples under pulsatile and oscillatory LAOE.

\subsection{Pulsatile flow}\label{sec_pulse}
\subsubsection{Newtonian fluid}
\begin{figure}
\centering
\includegraphics[width=0.5\textwidth]{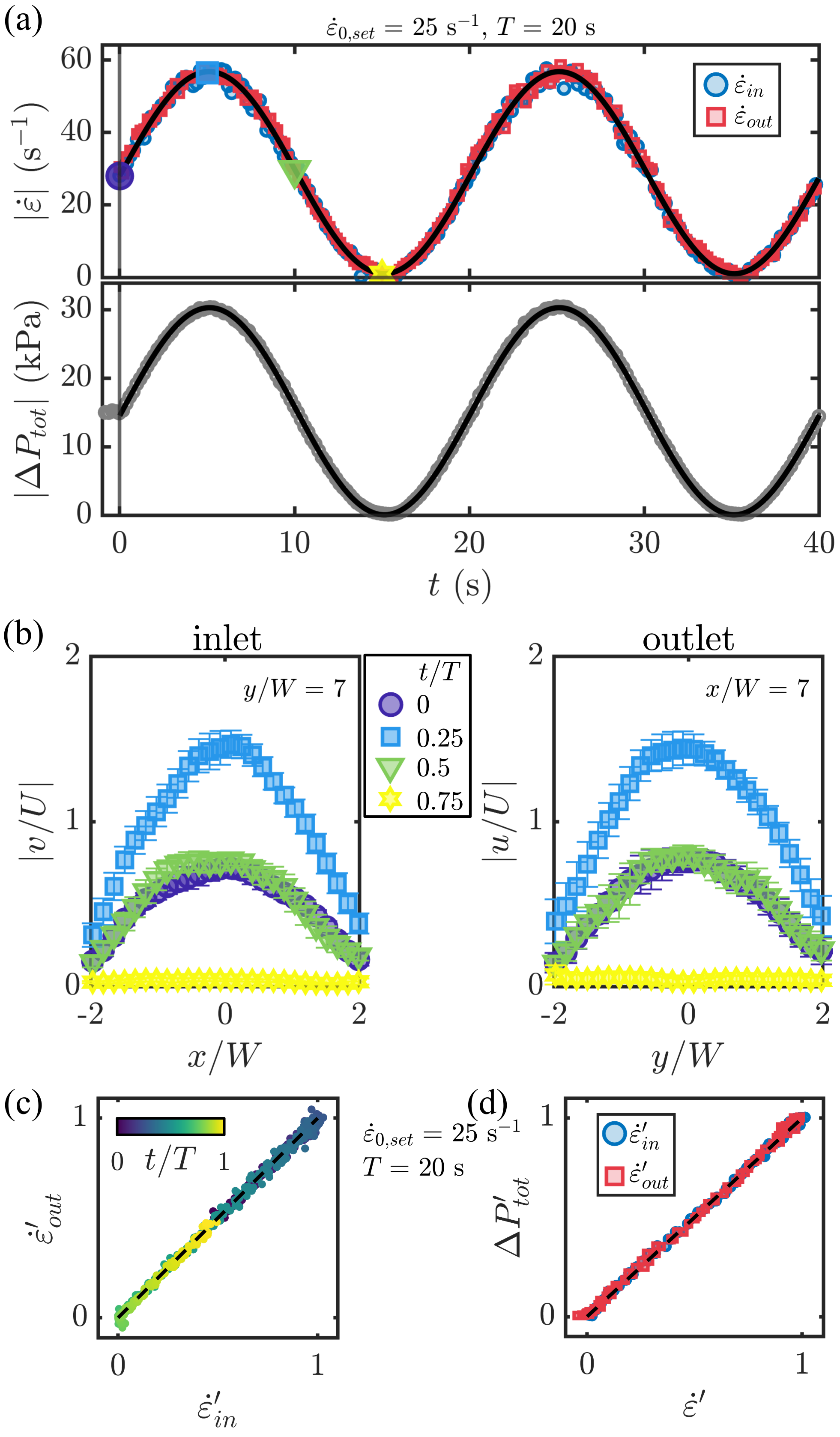}
\caption{Newtonian flow under pulsatile LAOE. (a) Representative raw measurements of the absolute values of the inlet \mbox{$\dot{\varepsilon}_{in}$} and outlet \mbox{$\dot{\varepsilon}_{out}$} extension rates (top) and the total pressure drop \mbox{$\Delta P_{tot}$} (bottom) for the Newtonian solvent at a set extension rate amplitude of \mbox{$\dot{\varepsilon}_{0,set}=\unit[25]{s^{-1}}$} and a period of \mbox{$T=\unit[20]{s}$}. Black lines indicate sinusoidal fits to the raw data. (b) Normalized profiles of the streamwise flow velocity at the inlet (right) and outlet (left) at various points during the cycle, corresponding to the four symbols highlighted in the extension rate evolution in (a). Error bars represent the standard deviation of phase-averaged data over multiple cycles. (c) Representative Lissajous curve of the normalized outlet extension rate \mbox{$\dot{\varepsilon}_{out}^\prime$} as a function of the normalized inlet extension rate \mbox{$\dot{\varepsilon}_{in}^\prime$}. (d) Representative Lissajous curve of the normalized total pressure drop, \mbox{$\Delta P_{tot}^\prime$}, as a function of the normalized extension rates, \mbox{$\dot{\varepsilon}_{in}^\prime$} and \mbox{$\dot{\varepsilon}_{out}^\prime$}, in the inlet and outlet directions. Data in (c) and (d) are shown for the same set extension rate amplitude and period as in (a). Black dashed lines in (c) and (d) correspond to slopes of 1.}
\label{FIG_PulseNewton}
\end{figure}

We begin by describing the flow of the Newtonian reference fluid under pulsatile driving conditions. Figure~\ref{FIG_PulseNewton}(a) illustrates the modulus of the extension rates \mbox{$\dot{\varepsilon}_{in}$} and \mbox{$\dot{\varepsilon}_{out}$} along with the total pressure drop \mbox{$\Delta P_{tot}$} at \mbox{$\dot{\varepsilon}_{0,set}=\unit[25]{s^{-1}}$} and \mbox{$T=\unit[20]{s}$}, shown over two periods. Similar to the steady-flow measurements, the extension rates at the compression axis \mbox{$\dot{\varepsilon}_{in}(t)=\partial v(t)/\partial y$} and extension axis \mbox{$\dot{\varepsilon}_{out}(t)=\partial u(t)/\partial x$} directions are calculated by averaging the velocity gradients over the relevant spatial domains for each image pair obtained during time-dependent PIV analysis without applying time-averaging.

Note that flow modifications in polymeric systems along the outlet direction can alter the local flow field, leading to a reduced local outlet strain rate \mbox{$\dot{\varepsilon}_{out}(t)$} at \mbox{$y=0$} along the extension axis (see Fig.~\ref{FIG_SteadyFlow}(d)). Consequently, the temporal outlet extension rate is not fitted to a sinusoidal model. Instead, \mbox{$\dot{\varepsilon}_{out}(t)$} is treated as an experimental measurement outcome, representing the fluid's response to the imposed inlet strain rate \mbox{$\dot{\varepsilon}_{in}(t)$}.

\paragraph{Flow field characterization}
As shown in Fig.~\ref{FIG_PulseNewton}(a), we find that \mbox{$\dot{\varepsilon}_{out}(t)=-\dot{\varepsilon}_{in}(t)$} for the Newtonian fluid under pulsatile driving, with both signals closely following the set sine signal. During the cycle, homogeneous velocity profiles are observed across the fluid inlet and outlet as long as \mbox{$U > 0$}, as representatively shown in Fig.~\ref{FIG_PulseNewton}(b) for various \mbox{$t/T$}. Furthermore, velocity profiles at equivalent temporal extension rates during the cycle, such as at \mbox{$t/T=0$} and \mbox{$t/T=0.5$}, exhibit identical values within experimental error, in agreement with the velocity profiles observed under steady flow conditions (see Fig.~\ref{FIG_SteadyFlow}(d)).

To further analyze the flow, we phase-average the extension rate profiles and plot the Lissajous figure of the normalized strain rate in the outlet direction \mbox{$\dot{\varepsilon}_{out}^\prime$} as a function of the normalized inlet extension rate \mbox{$\dot{\varepsilon}_{in}^\prime$} as shown for \mbox{$\dot{\varepsilon}_{0,set}=\unit[25]{s^{-1}}$} and \mbox{$T=\unit[20]{s}$} in Fig.~\ref{FIG_PulseNewton}(c). Similarly, we plot the normalized total pressure drop as a function of the normalized strain rates, shown in Fig.~\ref{FIG_PulseNewton}(d). Both Lissajous figures demonstrate that the response of the Newtonian fluid forms a straight line, indicating a linear system response with no phase shift between the signals. This linear behavior is consistent across all tested extension rate amplitudes \mbox{$\dot{\varepsilon}_{0,set}=\unit[0.5-25]{s^{-1}}$} and periods \mbox{$T=\unit[1-20]{s}$} as shown in Fig.~S5(a,b) in the Supplementary Material.

Overall, the results shown in Fig.~\ref{FIG_PulseNewton} and Fig.~S5 demonstrate the linear response of the Newtonian reference fluid under pulsatile LAOE, as evidenced by the straight-line Lissajous curves of \mbox{$\dot{\varepsilon}_{out}^\prime$} versus \mbox{$\dot{\varepsilon}_{in}^\prime$} and \mbox{$\Delta P_{tot}^\prime$} versus \mbox{$\dot{\varepsilon}^\prime$} over a broad range of applied strain rate amplitudes (\mbox{$\dot{\varepsilon}_{0,set}=\unit[0.5-25]{s^{-1}}$}) and pulsation periods (\mbox{$T=\unit[1-20]{s}$}). Similar to measurements under steady flow conditions, the flow field inside the OSCER remains symmetric during the time-dependent measurements, with homogeneous velocity profiles observed across the fluid inlet and outlet, both upstream and downstream of the stagnation point. 

\subsubsection{Polymer solution}
\begin{figure*}
\centering
\includegraphics[width=\textwidth]{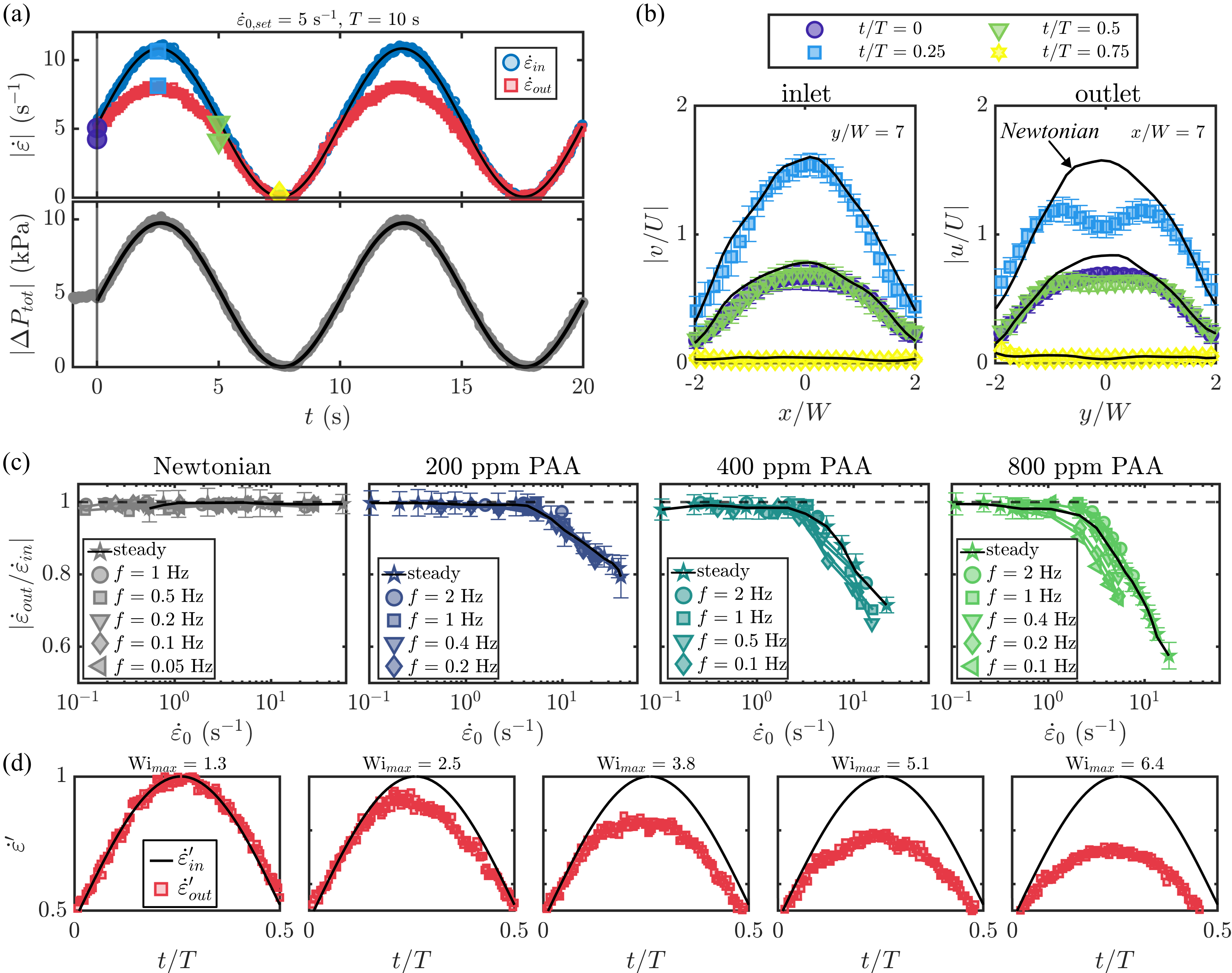}
\caption{Non-Newtonian flow under pulsatile LAOE. (a) Representative raw measurements of the absolute values of the inlet \mbox{$\dot{\varepsilon}_{in}$} and outlet \mbox{$\dot{\varepsilon}_{out}$} extension rates (top) and the total pressure drop \mbox{$\Delta P_{tot}$} (bottom) for the 800~ppm PAA sample, using a set extension rate amplitude of \mbox{$\dot{\varepsilon}_{0,set}=\unit[5]{s^{-1}}$} and a period of \mbox{$T=\unit[10]{s}$}. Solid black lines represent sinusoidal fits to the raw data. (b) Normalized profiles of the streamwise flow velocity at the inlet (right) and outlet (left) at various points during the cycle, as indicated by the four large symbols in the extension rate profiles in (a). Black lines in (b) show the experimentally measured velocity profiles for the Newtonian fluid at the same \mbox{$t/T$} as for the polymer solution. Error bars in (b) represent the standard deviation from phase averaging over multiple cycles. (c) Ratios between the measured maximum elongation rates in the outflow and inflow directions. Data are derived at the maximum temporal strain rate at \mbox{$t/T=0.25$} and are shown as a function of the inlet extension rate amplitude \mbox{$\dot{\varepsilon}_{0}$} for various pulsation frequencies. For the steady flow condition (highlighted by black lines), data are shown as the ratio between the steady outlet and inlet velocity gradients. (d) Zoom-in on the normalized outlet rate \mbox{$\dot{\varepsilon}_{out}^\prime$} during one half-cycle for various \mbox{$\text{Wi}_{max}$} for the 800~ppm PAA sample at \mbox{$\text{De}=0.12$}. Black lines correspond to fits of the normalized inlet strain rate profiles \mbox{$\dot{\varepsilon}_{in}^\prime$}.}
\label{FIG_PulseNonNewton}
\end{figure*}

\paragraph{Flow field characterization}
We proceed to examine the flow field of the viscoelastic PAA solutions under pulsatile LAOE. Figure~\ref{FIG_PulseNonNewton}(a) shows a representative measurement of the modulus of the average extension rates \mbox{$\dot{\varepsilon}_{in}$} and \mbox{$\dot{\varepsilon}_{out}$}, as well as the total pressure drop \mbox{$\Delta P_{tot}$} for the 800~ppm PAA sample at \mbox{$\dot{\varepsilon}_{0,set}=\unit[5]{s^{-1}}$} and \mbox{$T=\unit[10]{s}$} (\mbox{$\text{Wi}_{max}=6.4$} and \mbox{$\text{De}=0.06$}) over two periods. We observe that the inlet strain rate and the total pressure drop follow the sinusoidal signal imposed by the pumps. However, in contrast to the Newtonian case, a significant difference is found between the \mbox{$\dot{\varepsilon}_{in}$} and \mbox{$\dot{\varepsilon}_{out}$} profiles during the cycle. While \mbox{$\dot{\varepsilon}_{out}$} follows \mbox{$\dot{\varepsilon}_{in}$} at low strain rates (\mbox{$\lvert \dot{\varepsilon}_{in}(t)\rvert \lesssim \unit[4]{s^{-1}}$}), the measured average strain rate along the outflow axis is significantly smaller than the average strain rate along the inflow axis when the temporal extension rate increase during the cycle (\mbox{$\lvert \dot{\varepsilon}_{in}(t)\rvert \gtrsim \unit[4]{s^{-1}}$}).

The normalized profiles of the streamwise flow velocity at the inlet agree with those of the Newtonian fluid, as shown in Fig.~\ref{FIG_PulseNonNewton}(b). At the outlet axis, we observe a flattening of the central peak of the velocity profile and, eventually, the emergence of a local minimum at \mbox{$y/W=0$} along the stretching axis at large temporal strain rates, specifically at \mbox{$t/T=0.25$}. This observation is similar to the behavior under steady flow conditions (Fig.~\ref{FIG_SteadyFlow}(d)), where the flow field modification compared to the Newtonian case is associated with the localized polymer stretching downstream along the outlet centerline~\cite{Lyazid1980, Gardner1982, Dunlap1987, Harlen1990, Haward2010, Haward2012d}. We quantify the magnitude of this effect by plotting the modulus of the outlet to inlet extension rate ratio \mbox{$\lvert \dot{\varepsilon}_{out}/ \dot{\varepsilon}_{in}\rvert$} at \mbox{$t/T=0.25$}.

Figure~\ref{FIG_PulseNonNewton}(c) shows \mbox{$\lvert \dot{\varepsilon}_{out}/ \dot{\varepsilon}_{in}\rvert$} as a function of the inlet extension rate amplitude \mbox{$\dot{\varepsilon}_{0}$} and the pulsation frequency \mbox{$f$} for all four investigated samples. Additionally, we plot the ratio \mbox{$\lvert \dot{\varepsilon}_{out}/ \dot{\varepsilon}_{in}\rvert$} under steady flow conditions (Fig.~\ref{FIG_SteadyFlow}(g)) with \mbox{$\dot{\varepsilon}_{0}=\dot{\varepsilon}_{set}$}. For the Newtonian reference fluid, \mbox{$\lvert \dot{\varepsilon}_{out}/ \dot{\varepsilon}_{in}\rvert = 1$} under both pulsatile LAOE and steady flow conditions. For the polymer solutions, the ratio \mbox{$\lvert \dot{\varepsilon}_{out}/ \dot{\varepsilon}_{in}\rvert$} begins to decrease above a critical strain rate amplitude, independent of the applied frequency. The magnitude of this decrease increases with polymer concentration, similar to the decrease observed under steady flow conditions at \mbox{$\text{Wi} \geq 1$} (see Fig.~\ref{FIG_SteadyFlow}(g)).

The ratio of the average strain rate along the extension axis to that along the compression axis could also serve as an indicator of strain hardening. For Newtonian fluids or polymer samples at \mbox{$\text{Wi} \leq 1$}, strain hardening is not expected, resulting in \mbox{$\lvert \dot{\varepsilon}_{out}/\dot{\varepsilon}_{in}\rvert = 1$}. When strain hardening occurs, the fluid's extensional viscosity along the outlet direction increases locally, leading to \mbox{$\lvert \dot{\varepsilon}_{out}/\dot{\varepsilon}_{in}\rvert < 1$}. If the fluid transitions to a solid-like elastic state during extension, \mbox{$\lvert \dot{\varepsilon}_{out}/\dot{\varepsilon}_{in}\rvert \rightarrow 0$}, indicating plug-like motion along the extension axis.

For the polymeric sample used in this study, Fig.~\ref{FIG_PulseNonNewton}(d) shows how the outlet extension rate begins to deviate from the inlet rate as \mbox{$\text{Wi}_{max}$} is increased during the first half of the pulsation cycle. The black line corresponds to a sinusoidal fit of the inlet extension rate. At \mbox{$\text{Wi}_{max} = 1.3$}, the two strain rate profiles approximately overlap, indicating \mbox{$\lvert \dot{\varepsilon}_{out}/ \dot{\varepsilon}_{in}\rvert \approx 1$}. However, at \mbox{$\text{Wi}_{max} = 2.5$}, the outlet extension rate drops below the inlet rate at \mbox{$t/T \approx 0.125$} and remains lower until \mbox{$t/T \approx 0.5$}. The onset of this deviation shifts to smaller \mbox{$t/T$} as the Weissenberg number increases, while the time during the cycle when the outlet strain rate matches the inlet strain rate again shifts to larger \mbox{$t/T$}.

\begin{figure*}
\centering
\includegraphics[width=\textwidth]{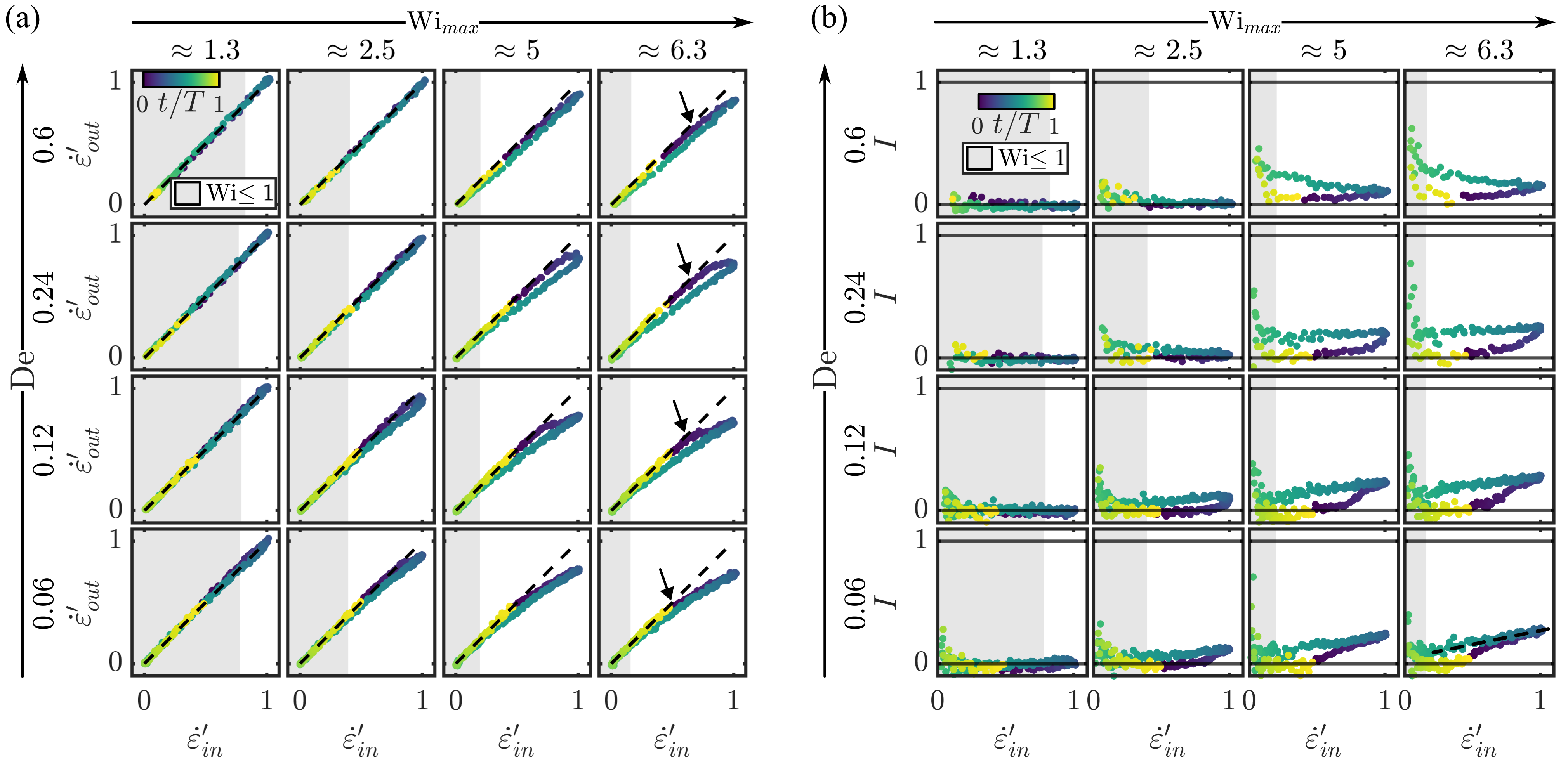}
\caption{Flow field characterization of a non-Newtonian fluid under pulsatile LAOE. Data is representatively shown for the 800~ppm PAA sample. (a) Lissajous curves of the normalized strain rate in the outlet direction \mbox{$\dot{\varepsilon}_{out}^\prime= \lvert \dot{\varepsilon}_{out}(t)/\dot{\varepsilon}_{max} \rvert$} as a function of the normalized inlet extension rate \mbox{$\dot{\varepsilon}_{in}^\prime= \lvert \dot{\varepsilon}_{in}(t)/\dot{\varepsilon}_{max} \rvert$} (with \mbox{$\dot{\varepsilon}_{max}=\dot{\varepsilon}_{0}+\lvert \dot{\varepsilon}_{off} \rvert$}) for various \mbox{$\text{De}$} and \mbox{$\text{Wi}_{max}$}. Black dashed lines correspond to slopes of 1, indicating the Newtonian case. The black arrows indicate the onset of the deviation from Newtonian behavior during the growth of the sinusoidal signal at \mbox{$\text{Wi}_{max} \approx 6.3$}. (b) Strain-hardening index \mbox{$I=1-\dot{\varepsilon}_{out}^\prime/\dot{\varepsilon}_{in}^\prime$} during the pulsation cycle. The \mbox{$\text{Wi}_{max}$} values above the panel correspond to approximate values at similar maximum Weissenberg numbers averaged over the shown \mbox{$\text{De}$} range. The dashed black line in the bottom left of panel (b) indicates a linear increase of \mbox{$I$}. Gray areas highlight the periods when \mbox{$\text{Wi}(t) \leq 1$} during the cycle, calculated based on the inlet strain rate.}
\label{FIG_PulseNonNewtonPIV}
\end{figure*}

In an effort to further explore the differences between the inlet and outlet extension rates during pulsatile LAOE, we plot the Lissajous figures of the normalized strain rate in the outlet direction \mbox{$\dot{\varepsilon}_{out}^\prime(t)$} as a function of the normalized inlet extension rate \mbox{$\dot{\varepsilon}_{in}^\prime(t)$} during the cycle in Fig.~\ref{FIG_PulseNonNewtonPIV}(a) for various \mbox{$\text{De}$} and \mbox{$\text{Wi}_{max}$}. At low Weissenberg numbers, \textit{i.e.}, \mbox{$\text{Wi}_{max} \lesssim 1.3$}, the response of the polymeric sample follows a straight line, similar to the Newtonian case, as indicated by the black dashed line. This behavior is independent of \mbox{$\text{De}$} within the investigated regime. As \mbox{$\text{Wi}_{max}$} increases while keeping the Deborah number constant, \textit{e.g.}, \mbox{$\text{De} = 0.06$}, we observe a sublinear increase in \mbox{$\dot{\varepsilon}_{out}^\prime$} as \mbox{$\dot{\varepsilon}_{in}^\prime$} increases during the cycle. While the data for \mbox{$\dot{\varepsilon}_{out}^\prime$} during the decreasing phase of \mbox{$\dot{\varepsilon}_{in}^\prime(t)$} closely follows the increasing part of the sinusoidal \mbox{$\dot{\varepsilon}_{in}^\prime(t)$} modulation at \mbox{$\text{De} = 0.06$}, a hysteresis loop emerges in the Lissajous curve of \mbox{$\dot{\varepsilon}_{out}^\prime$} versus \mbox{$\dot{\varepsilon}_{in}^\prime$} at \mbox{$\text{De} = 0.12$} and the data follows a different path during the decreasing part of the cycle. Moreover, the onset of the deviation from Newtonian behavior during the growth of the sinusoidal signal shifts to larger \mbox{$t/T$} during the cycle with increasing Deborah number at a fixed maximum Weissenberg number, as indicated by the arrows in Fig.~\ref{FIG_PulseNonNewtonPIV}(a) for \mbox{$\text{Wi}_{max} \approx 6.3$}. It is important to note that this deviation occurs for temporal Weissenberg numbers \mbox{$\text{Wi}(t) > 1$}, where \mbox{$\text{Wi}\leq 1$} is highlighted by the gray area in Fig.~\ref{FIG_PulseNonNewtonPIV}(a). For the 800~ppm PAA sample, we observe pronounced hysteresis loops for intermediate Deborah numbers \mbox{$0.06 < \text{De} < 0.6$} at \mbox{$\text{Wi}_{max} \gtrsim 2.5$}. In contrast, the hysteresis loop is less pronounced at both low (\mbox{$\text{De} = 0.06$}) and high (\mbox{$\text{De} = 0.6$}) Deborah numbers, even when \mbox{$\text{Wi}_{max} \geq 5$}.

To understand the temporal LAOE response of the polymer solution and to highlight the deviation from Newtonian flow during pulsatile LAOE, we introduce a strain-hardening index \mbox{$I=1-\dot{\varepsilon}_{out}^\prime/\dot{\varepsilon}_{in}^\prime$}. For an ideal Newtonian liquid, \mbox{$\dot{\varepsilon}_{out}(t)=\dot{\varepsilon}_{in}(t)$}, and hence, \mbox{$I=0$}. Indeed, we observe \mbox{$I\approx0$} for the Newtonian reference fluid under pulsatile LAOE, independent of the applied period \mbox{$T$} and amplitude \mbox{$\dot{\varepsilon}_{0,set}$}, as shown in Fig.~S5(c) in the Supplementary Material. For an ideal elastic solid, \mbox{$I=1$}, and for a viscoelastic fluid, we expect \mbox{$I$} to vary between \mbox{$0\leq I(t)\leq1$} during the cycle, depending on \mbox{$\text{De}$} and \mbox{$\text{Wi}_{max}$}. Figure~\ref{FIG_PulseNonNewtonPIV}(b) shows the evolution of the strain-hardening index during pulsatile LAOE for the 800~ppm PAA sample at various \mbox{$\text{De}$} and \mbox{$\text{Wi}_{max}$}. Similar to the observations in Fig.~\ref{FIG_PulseNonNewtonPIV}(a), we find that the polymeric sample follows Newtonian behavior with \mbox{$I\approx0$} at \mbox{$\text{Wi}_{max}\lesssim 1.3$}, regardless of the applied driving frequency (\mbox{$\text{De}$}). For low \mbox{$\text{De}$}, we observe an increase in the \mbox{$I$}-curve during the growing part of the cycle as the sample exhibits strain-hardening along the outlet direction with increasing maximum Weissenberg number. After reaching the maximum strain rate (\mbox{$\dot{\varepsilon}_{in}^\prime=1$}), the strain-hardening index decreases almost linearly as the inlet strain rate decreases, as indicated by the dashed black line for \mbox{$\text{De}=0.06$} and \mbox{$\text{Wi}_{max}=6.3$} in Fig.~\ref{FIG_PulseNonNewtonPIV}(b). For intermediate Deborah numbers \mbox{$0.06<\text{De}<0.6$} and Weissenberg numbers \mbox{$\text{Wi}_{max}\gtrsim2.5$}, a clear hysteresis loop emerges in the strain-hardening curves.

Note that the strain-hardening index data during pulsatile LAOE starts to scatter when the inlet strain rate \mbox{$\dot{\varepsilon}_{in}(t)$} approaches zero during the cycle. However, this apparent deviation from the Newtonian response is merely due to small fluctuations in the measured velocities in both the outlet and inlet directions when the flow reaches the minimum of the imposed sine signal.

We observe qualitatively similar behavior in the Lissajous curve of \mbox{$\dot{\varepsilon}_{out}^\prime$} versus \mbox{$\dot{\varepsilon}_{in}^\prime$} and the strain-hardening index for the 200~ppm and 400~ppm PAA samples, as shown in Fig.~S6 in the Supplementary Material. However, the magnitude of the hysteresis loops in the Lissajous curve at intermediate Deborah numbers is less pronounced for the lower-concentration polymer solutions compared to the 800~ppm PAA sample.

Taken together, the strain-hardening index allows us to visually determine when the flow of a viscoelastic fluid deviates from the Newtonian response during LAOE. We will use \mbox{$I$} in Sec.~\ref{sec_compare} to relate the onset of strain hardening with the emergence of elastic stresses, characterized by an increase in the excess pressure drop \mbox{$\Delta P_{ex}$}, in the polymeric systems.

\begin{figure*}
\centering
\includegraphics[width=\textwidth]{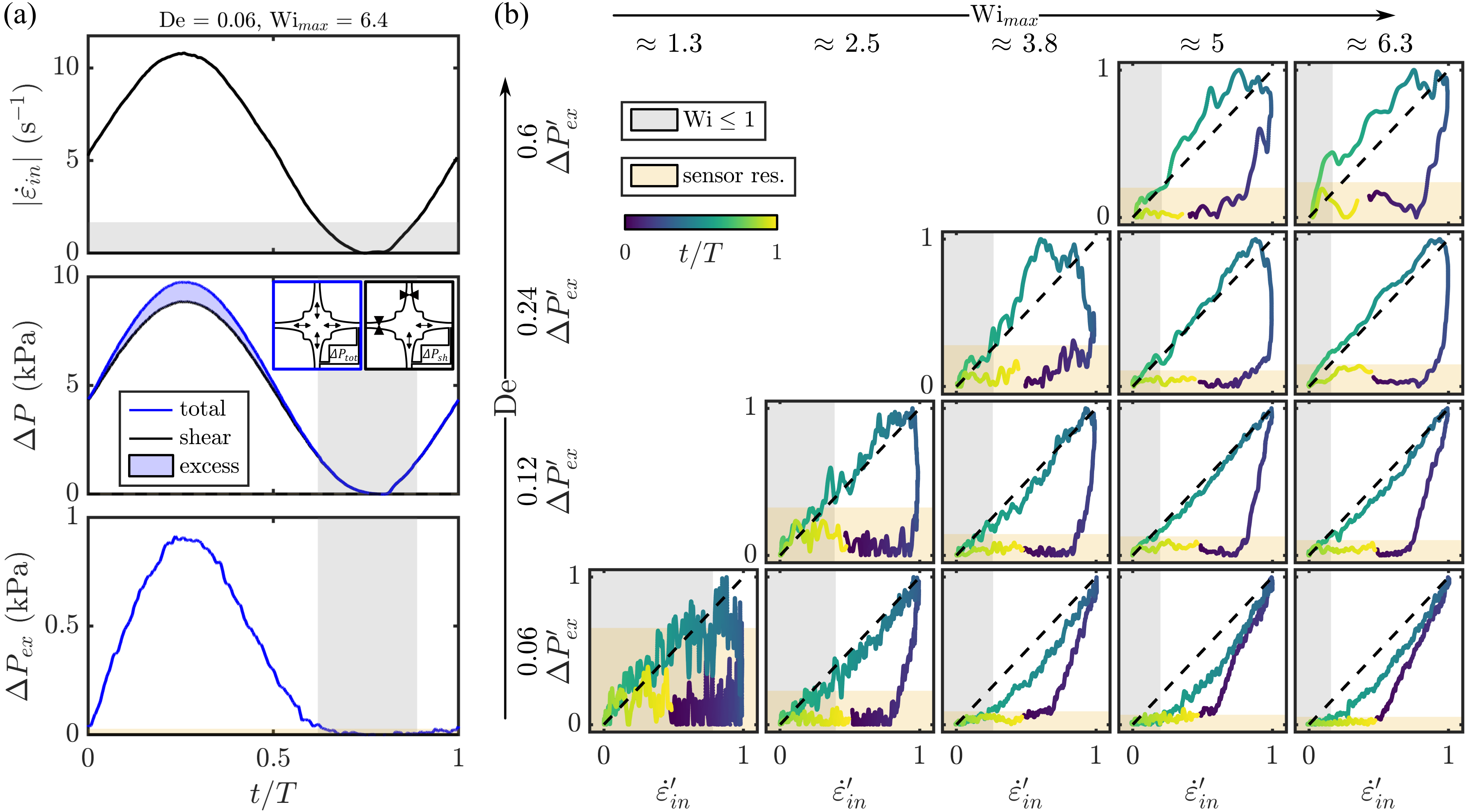}
\caption{Excess pressure during pulsatile LAOE. (a) Representative measurement for the 800~ppm PAA sample showing the absolute inlet extension rate (top), total and shear pressure drops (middle), and excess pressure drop (bottom) across the normalized pulsation cycle. Data corresponds to \mbox{$\text{De}=0.06$} and \mbox{$\text{Wi}_{max}=6.4$} (\mbox{$\dot{\varepsilon}_{0,set}=\unit[5]{s^{-1}}$} and  \mbox{$T=\unit[10]{s}$}). (b) Lissajous curves of the normalized excess pressure drop, \mbox{$\Delta P_{ex}^\prime$}, versus the normalized inlet extension rate, \mbox{$\dot{\varepsilon}_{in}^\prime$}, for various \mbox{$\text{De}$} and \mbox{$\text{Wi}_{max}$}. Black dashed lines indicate a slope of 1, representing the Newtonian case. Gray areas denote regions where \mbox{$\text{Wi}(t)\leq1$} during the cycle, and yellow areas indicate the lower detection limit of the pressure transducer. The \mbox{$\text{Wi}_{max}$} values above the panel correspond to approximate values at similar maximum Weissenberg numbers, averaged over the shown \mbox{$\text{De}$} range.}
\label{FIG_PulseExcessP}
\end{figure*}

\paragraph{Excess pressure drop}
In addition to measuring the flow field, we simultaneously measure the pressure drop during pulsatile LAOE, as described in Sec.~\ref{sec_pressuremeasure}. Complementary to the total pressure drop (see Fig.~S7 in the Supplementary Material), we calculate the excess pressure drop \mbox{$\Delta P_{ex}$} to evaluate the fluid’s elastic stress response. Figure~\ref{FIG_PulseExcessP}(a) presents a representative measurement of the total pressure drop, shear pressure drop, and excess pressure drop for the 800~ppm PAA sample at \mbox{$\text{De}=0.06$} and \mbox{$\text{Wi}_{max}=6.4$}.

The total and shear pressure drops follow the imposed sinusoidal profile during the pulsation. However, the total pressure drop is significantly larger than the shear pressure drop, particularly during the first half of the period. This results in a pronounced excess pressure drop \mbox{$\Delta P_{ex}$} for \mbox{$0 < t/T \lesssim 0.6$}, as shown in the bottom panel of Fig.~\ref{FIG_PulseExcessP}(a). For \mbox{$0.6 \lesssim t/T \lesssim 0.9$}, the excess pressure drop decreases to zero as the temporal Weissenberg number \mbox{$\text{Wi}(t)=\lambda \dot{\varepsilon}_{in}(t)$} falls below 1 during the cycle, as indicated by the gray area in Fig.~\ref{FIG_PulseExcessP}.

Figure~\ref{FIG_PulseExcessP}(b) shows the relationship between the normalized excess pressure drop \mbox{$\Delta P_{ex}^\prime= \lvert \Delta P_{ex}(t)/\Delta P_{ex,max} \rvert$} and the normalized inlet extension rate \mbox{$\dot{\varepsilon}_{in}^\prime$} for various \mbox{$\text{De}$} and \mbox{$\text{Wi}_{max}$}. For the specific data shown in (a) (\mbox{$\text{De}=0.06$} and \mbox{$\text{Wi}_{max}\approx6.3$} in Fig.~\ref{FIG_PulseExcessP}(b)), the excess pressure drop increases linearly with the imposed strain rate at the beginning of the cycle. Once it reaches its maximum value at \mbox{$\dot{\varepsilon}_{in}^\prime=1$}, \mbox{$\Delta P_{ex}^\prime$} follows a concave decay until it drops below the detection limit of the pressure transducer as \mbox{$\text{Wi}(t)\leq1$}. In the latter part of the cycle, even as the strain rate rises again, the excess pressure drop does not recover but remains below the detection threshold. 

Decreasing \mbox{$\text{Wi}_{max}$} while keeping \mbox{$\text{De}=0.06$} constant results in a delayed initial increase of the excess pressure drop at the beginning of the cycle. This delay is characterized by a plateau at \mbox{$\Delta P_{ex}^\prime\approx0$} starting at \mbox{$\dot{\varepsilon}_{in}^\prime=0.5$}, which becomes more pronounced with decreasing \mbox{$\text{Wi}_{max}$}. Once \mbox{$\Delta P_{ex}^\prime$} begins to increase, its slope in the Lissajous plots \mbox{$\Delta P_{ex}^\prime(\dot{\varepsilon}_{in}^\prime)$} steepens as \mbox{$\text{Wi}_{max}$} decreases. After reaching the maximum excess pressure drop, the decay transitions from a convex to a more linear profile (indicated by the dashed black line in Fig.~\ref{FIG_PulseExcessP}(b)) as \mbox{$\text{Wi}_{max}$} decreases, approaching \mbox{$\Delta P_{ex}^\prime\approx0$} at \mbox{$\dot{\varepsilon}_{in}^\prime=0$}. This trend continues until the measured signals approach the lower detection limit of the pressure transducer, notably at \mbox{$\text{Wi}_{max}\approx1.3$} and \mbox{$\text{De}=0.06$}.

A similar behavior is observed when \mbox{$\text{Wi}_{max}$} is fixed, and \mbox{$\text{De}$} is increased. At \mbox{$\text{Wi}_{max}\approx6.3$} and \mbox{$\text{De}=0.12$}, the initial pressure increase is delayed compared to \mbox{$\text{De}=0.06$}, with the delay (plateau at \mbox{$\Delta P_{ex}^\prime\approx0$}) growing as \mbox{$\text{De}$} increases. Concurrently, the slope of \mbox{$\Delta P_{ex}^\prime(\dot{\varepsilon}_{in}^\prime)$} increases once the excess pressure drop rises above zero. The subsequent decay of \mbox{$\Delta P_{ex}^\prime$} transitions from a concave to a more linear profile as \mbox{$\text{De}$} increases. At higher \mbox{$\text{De}$}, specifically \mbox{$\text{De}\geq0.24$}, and moderate \mbox{$\text{Wi}_{max}\approx3.8$}, the magnitude of the hysteresis loop in the Lissajous curve increases further. Simultaneously, the decay of \mbox{$\Delta P_{ex}$} after reaching its maximum undergoes another transition, shifting from a linear to a convex shape that surpasses the diagonal, as shown by the dashed black line in Fig.~\ref{FIG_PulseExcessP}(b). 

Similar trends with growing hysteresis loops of excess pressure drop \mbox{$\Delta P_{ex}^\prime$} versus inlet strain rate \mbox{$\dot{\varepsilon}_{in}^\prime$} as a function of \mbox{$\text{De}$}, are also observed for the 200~ppm and 400~ppm PAA samples (see Fig.~S8 in the Supplementary Material). Note that plotting \mbox{$\Delta P_{ex}^\prime$} versus the outlet strain rate \mbox{$\dot{\varepsilon}_{out}^\prime$} does not significantly change the shape of the Lissajous curves (see Fig.~S9 in the Supplementary Material).

The peculiar shapes of the Lissajous curves of excess pressure drop versus inlet strain rate \mbox{$\Delta P_{ex}^\prime(\dot{\varepsilon}_{in}^\prime)$} resemble the banana-shaped loops previously reported in large-amplitude oscillatory uniaxial extension of elastomeric solids~\cite{Dessi2017}. Dessi~\textit{et al.}~\cite{Dessi2017} described nonlinear stress-strain responses in elastomers, attributed to strain-hardening of the system, and evidenced by Lissajous plots of measured stress responses versus uniaxial strain that exhibited convex or concave banana-shaped patterns. Similar nonlinear Lissajous curves of stress versus strain have been reported for soft polymeric networks and polymer melts~\cite{Rasmussen2008, Bejenariu2010}. 

In this study, we report the nonlinear viscoelastic response of dilute polymer solutions as a function of the strain rate, \textit{i.e.}, at the inlet \mbox{$\dot{\varepsilon}_{in}(t)$}, rather than the extensional strain \mbox{$\varepsilon(t)$} during the cycle. Similar to the characterization of nonlinear viscoelasticity in large-amplitude oscillatory shear, the curves shown in Fig.\ref{FIG_PulseExcessP}(b), depicting stress as a function of the imposed strain rate, can be referred to as viscous Lissajous curves. These are distinct from elastic Lissajous curves, representing stress as a function of strain~\cite{Ewoldt2008}.

In summary, these results highlight the unique viscoelastic responses of polymer solutions under pulsatile LAOE. The PIV measurements reveal deviations from straight-line Lissajous curves of \mbox{$\dot{\varepsilon}_{out}^\prime$} versus \mbox{$\dot{\varepsilon}_{in}^\prime$}, with pronounced hysteresis loops emerging at intermediate \mbox{$\text{De}$} and higher \mbox{$\text{Wi}_{max}$}. Combining this PIV analysis with excess pressure drop measurements reveals distinct triangular Lissajous curves in excess pressure drop versus inlet strain rate \mbox{$\Delta P_{ex}^\prime(\dot{\varepsilon}_{in}^\prime)$}, whose size and shape strongly depend on the applied \mbox{$\text{Wi}$} and \mbox{$\text{De}$}.

We note that measurements at \mbox{$\text{De}=0.6$} for the 800~ppm PAA sample in Fig.\ref{FIG_PulseExcessP} correspond to driving frequencies of \mbox{$f=\unit[1]{Hz}$}. At these frequencies, the syringe pumps, with a temporal resolution of approximately \mbox{$\unit[20]{ms}$}, produce only 50 discrete steps to approximate the sine wave. Combined with the large piston driving amplitudes to reach large maximum Weissenberg numbers, the pumps must move rapidly, making it increasingly challenging to generate a smooth sinusoidal signal. Consequently, pulsation frequencies of \mbox{$f\approx\unit[1]{Hz}$} represent the limit of the present experimental LAOE realization.

\subsection{Oscillatory flow}\label{sec_osc}
Besides the pulsatile LAOE mode, we further examine the response of the Newtonian reference fluid and the polymeric samples under pure oscillating flow conditions, applying a sinusoidal profile with \mbox{$\dot{\varepsilon}_{off,set}=0$} in Eq.\ref{eq_setsin}. In this oscillatory LAOE mode, the inlet and outlet channels, through which the fluid is injected and withdrawn, alternate between the \mbox{$x$} and \mbox{$y$} directions every half cycle, as detailed in Sec.~\ref{sec_timeMeasurements}. Consequently, the stretching axis in the OSCER flips by \mbox{$90^{\circ}$} every half period between the \mbox{$x$} and \mbox{$y$} axes. This alternation causes the strain rates in the inlet \mbox{$\dot{\varepsilon}_{in}(t)$} and outlet \mbox{$\dot{\varepsilon}_{out}(t)$} directions to switch every half period at \mbox{$t/T=0.5$} between the \mbox{$x$} and \mbox{$y$} axes. In contrast, during pulsatile LAOE and under steady flow conditions, the stretching axis remains fixed along the \mbox{$x$} axis. Under oscillatory LAOE, the fluid is effectively extended twice per imposed cycle, once along the \mbox{$x$} direction (for \mbox{$0<t/T<0.5$}) and once along the \mbox{$y$} direction (for \mbox{$0.5<t/T<1$}) (see Fig.~\ref{FIG_Setup}(c)). As a result, the effective period duration for oscillatory LAOE is \mbox{$T/2$} of the imposed sinusoidal signal \mbox{$\dot{\varepsilon}_{set}(t)$}. Consequently, the Deborah number for oscillatory LAOE measurements is calculated as \mbox{$\text{De}=2\lambda/T$}.

\subsubsection{Newtonian fluid}
\begin{figure*}
\centering
\includegraphics[width=\textwidth]{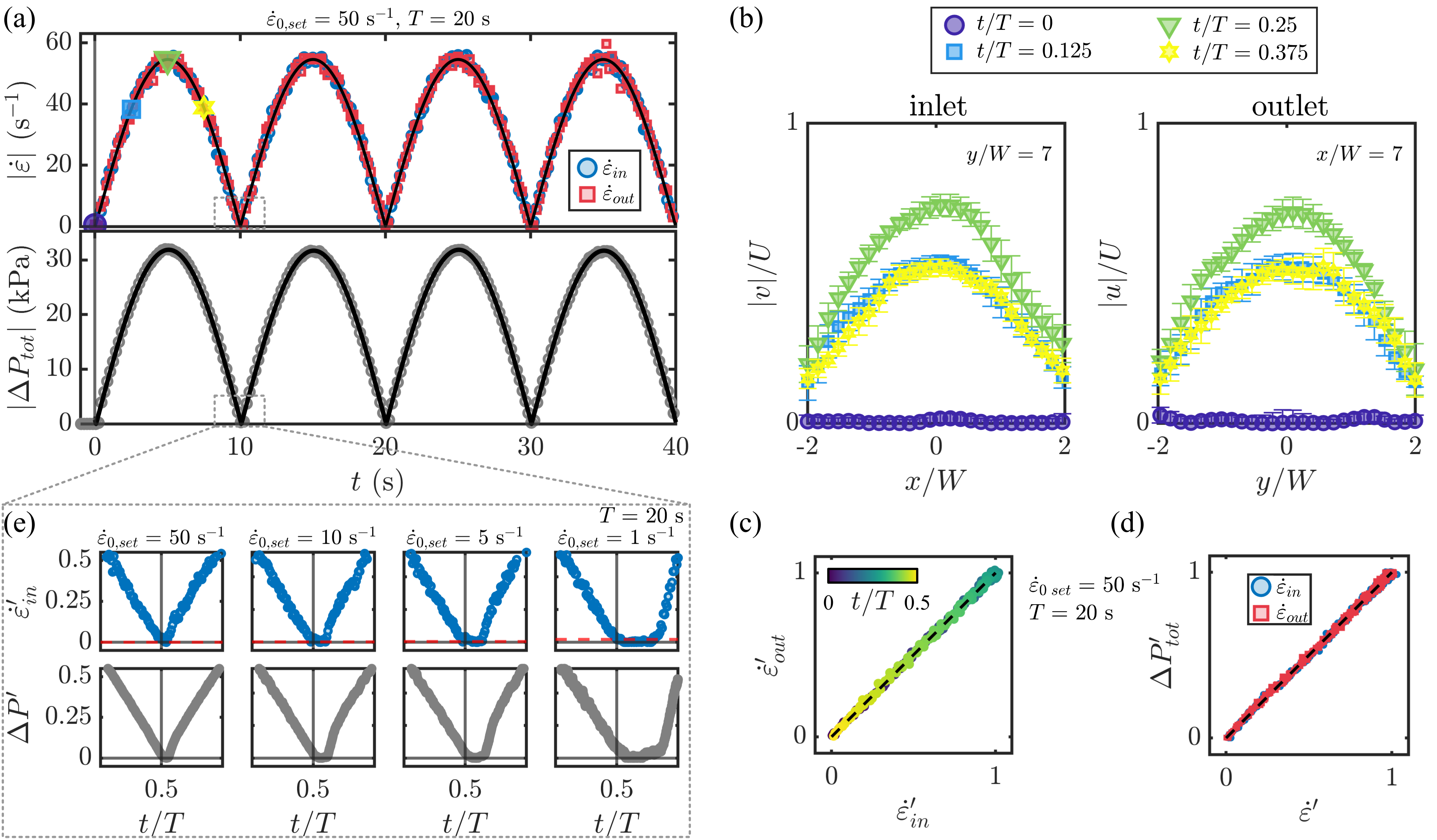}
\caption{Newtonian flow under oscillatory LAOE. (a) Representative raw measurements of the absolute values of the inlet \mbox{$\dot{\varepsilon}_{in}$} and outlet \mbox{$\dot{\varepsilon}_{out}$} extension rates (top) and total pressure drop \mbox{$\Delta P_{tot}$} (bottom) for the Newtonian solvent, using an extension rate amplitude of \mbox{$\dot{\varepsilon}_{0,set}=\unit[50]{s^{-1}}$} and a period of \mbox{$T=\unit[20]{s}$}. Black lines represent sinusoidal fits to the raw data. (b) Normalized profiles of the streamwise flow velocity at the inlet (right) and outlet (left) at various points during the cycle, as indicated by the four prominent symbols in the extension rate evolution shown in (a). Error bars in (b) represent the standard deviation from phase averaging over multiple cycles. (c) Representative Lissajous curve of the normalized outlet \mbox{$\dot{\varepsilon}_{out}^\prime$} versus the normalized inlet \mbox{$\dot{\varepsilon}_{in}^\prime$} extension rates. (d) Representative Lissajous curve of the normalized total pressure drop \mbox{$\Delta P_{tot}^\prime$} as a function of the normalized extension rates \mbox{$\dot{\varepsilon}_{in}^\prime$} and \mbox{$\dot{\varepsilon}_{out}^\prime$} in the inlet and outlet directions. Data in (c) and (d) are shown for the same extension rate amplitude and period as in (a). Black dashed lines in (c) and (d) correspond to slopes of 1. (e) Magnification of the normalized inlet extension rate \mbox{$\dot{\varepsilon}_{in}^\prime$} (top) and normalized total pressure drop \mbox{$\Delta P_{tot}^\prime$} (bottom) around \mbox{$t/T=0.5$} for various \mbox{$\dot{\varepsilon}_{0,set}$}. The red line in the top panel of (e) indicates the recommended pulse-free limit \mbox{$Q_{min}=\unit[0.16]{\upmu L/min}$}.}
\label{FIG_OscNewton}
\end{figure*}

Figure~\ref{FIG_OscNewton} summarizes the flow behavior of the Newtonian reference fluid under oscillatory LAOE. Similar to pulsatile LAOE, we observe \mbox{$\dot{\varepsilon}_{out}(t) = -\dot{\varepsilon}_{in}(t)$} for the Newtonian fluid, with both strain rates and the pressure signal following the magnitude of the imposed sinusoidal signal, as shown in Fig.~\ref{FIG_OscNewton}(a). Throughout the cycle, the velocity profiles across the fluid inlet and outlet show parabolic shapes, as illustrated in Fig.~\ref{FIG_OscNewton}(b) for various \mbox{$t/T$}. Velocity profiles at comparable temporal extension rates during the cycle, \textit{e.g.}, at \mbox{$t/T=0.125$} and \mbox{$t/T=0.375$}, overlap within the experimental error.

As in pulsatile LAOE, the Lissajous figure of the normalized strain rate in the outlet direction \mbox{$\dot{\varepsilon}_{out}^\prime(t)$} as a function of the normalized inlet extension rate \mbox{$\dot{\varepsilon}_{in}^\prime(t)$} (Fig.~\ref{FIG_OscNewton}(c)), and the Lissajous figure of the normalized total pressure drop \mbox{$\Delta P_{tot}^\prime$} as a function of the normalized extension rates (Fig.~\ref{FIG_OscNewton}(d)) both follow straight lines. This behavior indicates the linear response of the Newtonian fluid under oscillatory LOAE. This linearity is consistent across all investigated extension rate amplitudes (\mbox{$\dot{\varepsilon}_{0,set}=\unit[1-50]{s^{-1}}$}) and periods (\mbox{$T=\unit[1-20]{s}$}), as shown in Fig.~S10(a,b) in the Supplementary Material. For the Newtonian reference fluid subjected to oscillatory LAOE, \mbox{$I \approx 0$}, regardless of the applied period \mbox{$T$} and amplitude \mbox{$\dot{\varepsilon}_{0,set}$}, as shown in Fig.~S10(c). 

Reversing the pump direction and switching between injection \mbox{$Q_{in}$} and withdrawal \mbox{$Q_{out}$} operations at \mbox{$t/T=0.5$} can lead to deviations from the intended sinusoidal strain rate profiles. This effect is summarized in Figure~\ref{FIG_OscNewton}(e), which shows the normalized inlet strain rate \mbox{$\dot{\varepsilon}_{in}^\prime$} (top) and the measured total pressure drop \mbox{$\Delta P_{tot}^\prime$} (bottom) for a period of \mbox{$T=\unit[20]{s}$} and various set strain rate amplitudes \mbox{$\dot{\varepsilon}_{0,set}$}.

At high strain rate amplitudes, such as \mbox{$\dot{\varepsilon}_{0,set}=\unit[50]{s^{-1}}$}, both the inlet strain rate and pressure drop follow the expected profiles of the magnitude of the imposed sinusoidal signal at \mbox{$t/T=0.5$}. However, as the set strain rate amplitude decreases, a plateau appears in the inlet strain rate (\mbox{$\dot{\varepsilon}_{in}^\prime\approx0$}) around \mbox{$t/T\approx0.5$} when the pump reverses direction, indicating a momentary pause in the piston movement. The duration of this plateau progressively increases as the set strain rate amplitude decreases. When the piston's speed is reduced during oscillation, the pumps eventually approach the minimum flow rate they can sustain. For the specific pumps and \mbox{$\unit[1]{mL}$} syringes used, the recommended pulse-free limit is \mbox{$Q_{min}=\unit[0.16]{\upmu L/min}$} (\mbox{$\dot{\varepsilon}_{min}\approx\unit[0.01]{s^{-1}}$}), as indicated by the red lines in Figure~\ref{FIG_OscNewton}(e). This delay in the strain rate signal results in a corresponding plateau in the pressure signal for the same duration.

During pulsatile LAOE, the pumps also approach the minimum flow rate during the decreasing phase of the sine wave, leading to a small plateau in the inlet strain rate and pressure drop signals (see Fig.~\ref{FIG_PulseExcessP}(a) at \mbox{$t/T\approx0.75$}). Notably, we used a constant background flow with \mbox{$\dot{\varepsilon}_{off,set}=\dot{\varepsilon}_{0,set}$} for all pulsatile measurements conducted in this study. However, applying pulsatile LAOE with \mbox{$\dot{\varepsilon}_{off,set}>\dot{\varepsilon}_{0,set}$} would result in a strain rate signal where \mbox{$\dot{\varepsilon}(t)>0$} throughout the cycle, ensuring that the flow rate remains above the minimum operational limit of the syringe pumps. Selecting a constant background flow slightly exceeding the pulsation amplitude would therefore eliminate the plateau in the inlet strain rate and pressure profiles.

The emergence of the plateau and the deviation from the intended sinusoidal waveform at lower set strain rate amplitudes under oscillatory LAOE underscores a limitation of the oscillatory LAOE technique. This limitation is particularly apparent compared to pulsatile LAOE, where the pumps avoid reversing direction. Nonetheless, by simultaneously monitoring the time-dependent flow field and pressure drop, we can establish a clear relationship between the two signals.

\subsubsection{Polymer solution}
\begin{figure*}
\centering
\includegraphics[width=\textwidth]{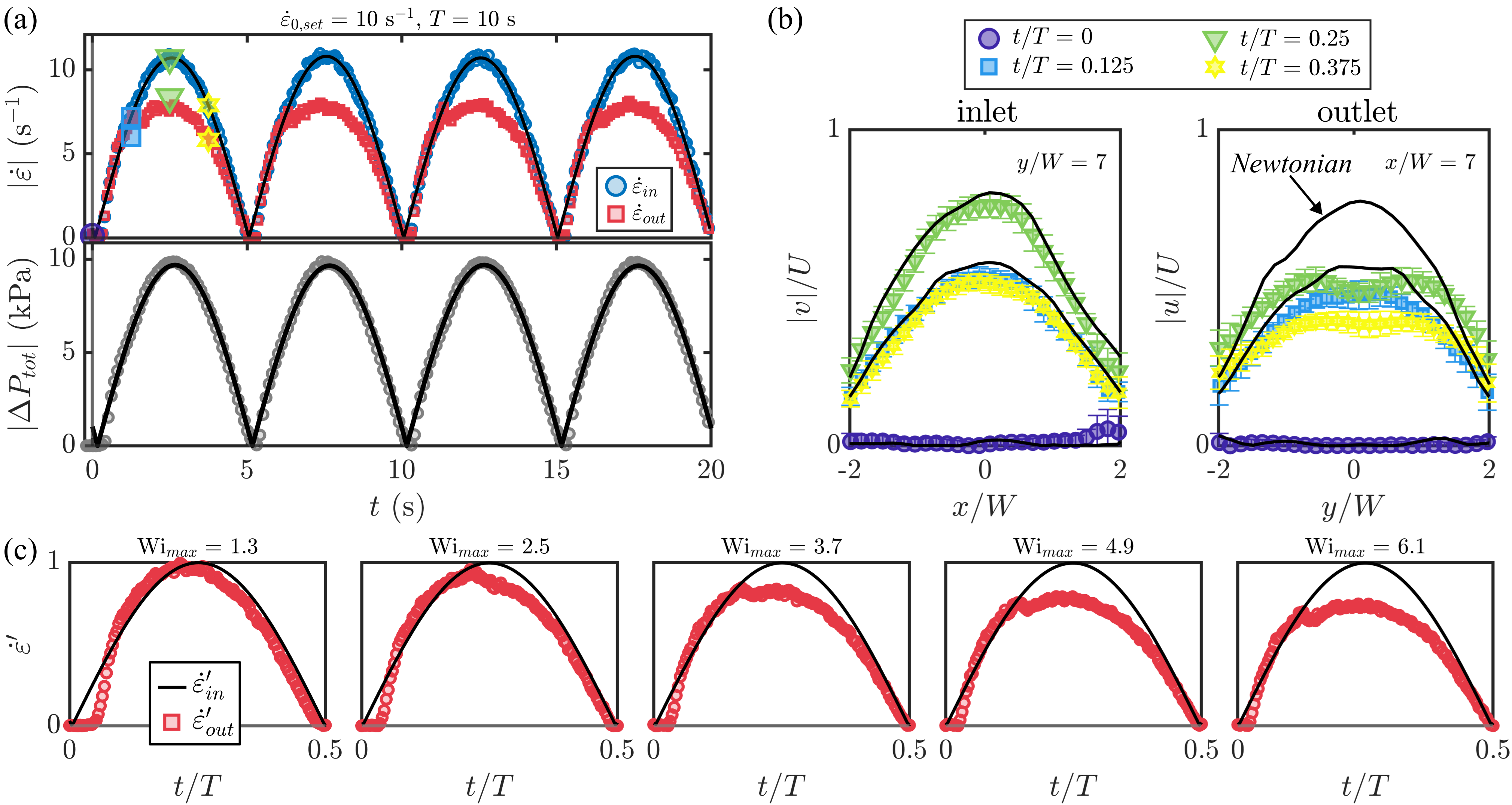}
\caption{Non-Newtonian flow under oscillatory LAOE. (a) Representative raw measurements of the absolute values of the inlet \mbox{$\dot{\varepsilon}_{in}$} and outlet \mbox{$\dot{\varepsilon}_{out}$} extension rates (top) and the total pressure drop \mbox{$\Delta P_{tot}$} (bottom) for the 800~ppm PAA sample using a set extension rate amplitude of \mbox{$\dot{\varepsilon}_{0,set}=\unit[10]{s^{-1}}$} and a period of \mbox{$T=\unit[10]{s}$}. Black lines represent sinusoidal fits to the raw data. (b) Normalized profiles of the streamwise flow velocity at the inlet (right) and outlet (left) at various points during the cycle, as indicated by the four prominent symbols in the extension rate evolution shown in (a). Black lines in (b) show the velocity profiles for the Newtonian fluid. Error bars in (b) represent the standard deviation from phase averaging over multiple cycles. (c) Zoomed-in view of the normalized outlet strain rate \mbox{$\dot{\varepsilon}_{out}^\prime$} during one half-cycle for various \mbox{$\text{Wi}_{max}$} values for the 800~ppm PAA sample at \mbox{$\text{De}=0.24$}. Black lines correspond to fits of the normalized inlet strain rate profiles \mbox{$\dot{\varepsilon}_{in}^\prime$}.}
\label{FIG_OscNonNewton}
\end{figure*}

\paragraph{Flow field characterization}
Figure~\ref{FIG_OscNonNewton}(a) presents a representative measurement of the average extension rates \mbox{$\dot{\varepsilon}_{in}$} and \mbox{$\dot{\varepsilon}_{out}$} along with the total pressure drop \mbox{$\Delta P_{tot}$} for the 800~ppm PAA sample at \mbox{$\dot{\varepsilon}_{0,set}=\unit[5]{s^{-1}}$} and \mbox{$T=\unit[10]{s}$} over two oscillation cycles. Similar to the Newtonian case, we observe that both the inlet strain rate and the overall pressure drop follow the modulus of the imposed sinusoidal signal, which corresponds to the pump modulation. However, unlike the Newtonian case, there is a notable difference between the \mbox{$\dot{\varepsilon}_{in}$} and \mbox{$\dot{\varepsilon}_{out}$} profiles throughout the cycle. While the two signals overlap at lower strain rates, specifically when \mbox{$\lvert \dot{\varepsilon}_{in}(t)\rvert \lesssim \unit[6]{s^{-1}}$}, the measured strain rate along the outflow axis is significantly smaller than that along the inflow axis when \mbox{$\lvert \dot{\varepsilon}_{in}(t)\rvert \gtrsim \unit[6]{s^{-1}}$}.

Similar to the behavior observed under pulsatile LAOE, the normalized streamwise flow velocity profiles at the inlet align with those of a Newtonian fluid, as shown in Fig.~\ref{FIG_OscNonNewton}(b). At the outlet, the central peak of the velocity profile flattens, and a local minimum emerges at \mbox{$y/W=0$} along the stretching axis when large temporal strain rates are applied at \mbox{$t/T=0.25$}, attributed to the localized stretching of the polymer downstream along the outlet centerline~\cite{Lyazid1980, Gardner1982, Dunlap1987, Harlen1990, Haward2010, Haward2012d}. Moreover, the outlet velocity profiles measured at similar inlet strain rates during the rising and falling phases of the sine wave, for example, at \mbox{$t/T=0.125$} and \mbox{$t/T=0.375$}, show significant differences. The profile at \mbox{$t/T=0.375$} retains a flattened shape with a minimum, while the profile at \mbox{$t/T=0.125$} is more parabolic. This suggests the presence of residual stresses in the polymer solution, which relax slowly compared to the driving frequency as the imposed strain rate decreases.

Figure~\ref{FIG_OscNonNewton}(c) illustrates this phenomenon for oscillatory LAOE, showing how the outlet extension rate begins to deviate from the inlet rate during the cycle as \mbox{$\text{Wi}_{max}$} increases at \mbox{$\text{De}=0.24$}. For all maximum Weissenberg numbers, the outlet strain rate profiles display asymmetric shapes on the rising and falling sides of the sine wave, compared to the fit of the inlet strain rate. However, it should be noted that the plateau in the strain rate profile at the beginning of the period (\mbox{$t/T=0$}) mainly results from experimental limitations when reversing the piston direction at the minimum operational limit of the syringe pumps, as discussed in the previous section and in Fig.~\ref{FIG_OscNewton}(e). At \mbox{$\text{Wi}_{max}=1.3$}, the two strain rate profiles nearly overlap once the initial plateau is surpassed. However, at \mbox{$\text{Wi}_{max}=2.5$}, the outlet extension rate falls below the inlet rate at \mbox{$t/T\approx0.225$}. This deviation starts at smaller \mbox{$t/T$} values as the maximum Weissenberg number increases, similar to the behavior observed under pulsatile LAOE. 

\begin{figure*}
\centering
\includegraphics[width=\textwidth]{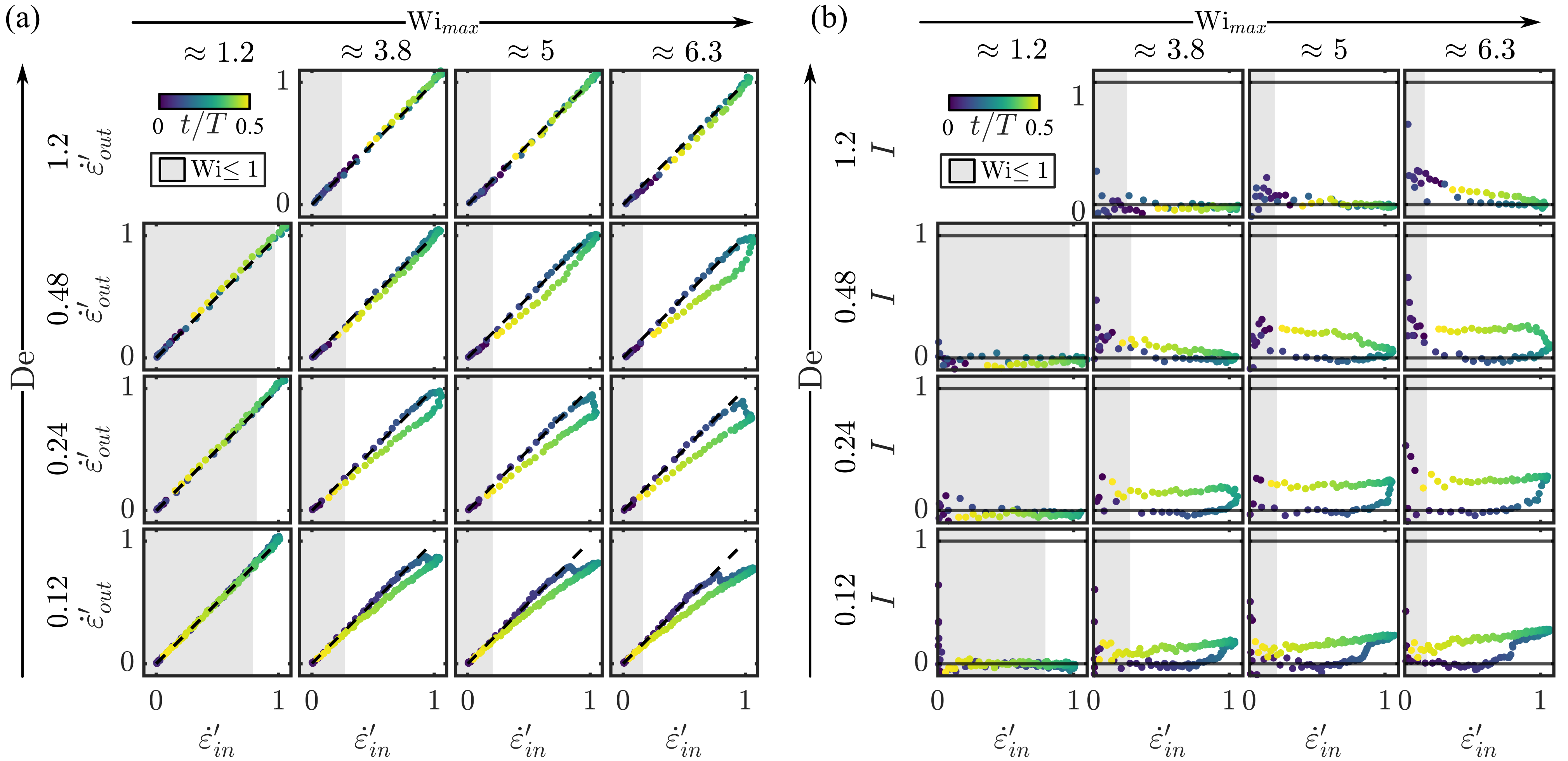}
\caption{Flow field characterization of a non-Newtonian fluid under oscillatory LAOE. Data are representatively shown for the 800~ppm PAA sample. (a) Lissajous curves of the normalized strain rate in the outlet direction \mbox{$\dot{\varepsilon}_{out}^\prime$}, as a function of the normalized inlet extension rate \mbox{$\dot{\varepsilon}_{in}^\prime$}, for various \mbox{$\text{De}$} and \mbox{$\text{Wi}_{max}$}. Black dashed lines correspond to slopes of 1, representing the Newtonian case. (b) Strain-hardening index \mbox{$I=1-\dot{\varepsilon}_{out}^\prime/\dot{\varepsilon}_{in}^\prime$}  during the cycle. The \mbox{$\text{Wi}_{max}$} values above the panel correspond to approximate values at similar maximum Weissenberg numbers, averaged over the shown \mbox{$\text{De}$} range. Gray areas represent regions where \mbox{$\text{Wi}(t)\leq 1$} during the cycle, calculated based on the inlet strain rate.}
\label{FIG_OscNonNewtonPIV}
\end{figure*}

In a manner similar to the analysis of both extension rates under pulsatile LAOE, we present the Lissajous figures for the normalized strain rate in the outlet direction \mbox{$\dot{\varepsilon}_{out}^\prime$} as a function of the normalized inlet extension rate \mbox{$\dot{\varepsilon}_{in}^\prime$} during the cycle. These results are shown for the 800~ppm PAA sample in Fig.~\ref{FIG_OscNonNewtonPIV}(a) for various \mbox{$\text{De}$} and \mbox{$\text{Wi}_{max}$}. At low Weissenberg numbers, specifically when \mbox{$\text{Wi}_{max}\lesssim 1.2$}, the data follows a straight line resembling the Newtonian case, as indicated by the black dashed line. This behavior is consistent regardless of the \mbox{$\text{De}$} value. Even with \mbox{$\text{De}= 1.2$}, the polymeric sample’s response remains nearly linear up to the highest Weissenberg number examined. As we increase \mbox{$\text{Wi}_{max}$} while keeping \mbox{$\text{De}< 1.2$}, the data initially shows a linear increase, aligning with the Newtonian reference line. However, at a critical strain rate, \mbox{$\dot{\varepsilon}_{out}^\prime$} starts to fall below \mbox{$\dot{\varepsilon}_{in}^\prime$}, and hysteresis loops appear in the Lissajous curve as the inlet strain rate decreases during the descending phase of the sine wave. This deviation occurs for temporal Weissenberg numbers \mbox{$\text{Wi}(t)>1$}, \textit{i.e.}, beyond the highlighted the gray areas in Fig.~\ref{FIG_OscNonNewtonPIV}(a).

Figure~\ref{FIG_OscNonNewtonPIV}(b) shows the evolution of the strain-hardening index \mbox{$I$} during oscillatory LAOE for the 800~ppm PAA sample, for various \mbox{$\text{De}$} and \mbox{$\text{Wi}_{max}$}. When \mbox{$\text{Wi}_{max} \lesssim 1.2$}, the polymeric sample exhibits Newtonian behavior with \mbox{$I \approx 0$}, independent of the applied \mbox{$\text{De}$}. As \mbox{$\text{Wi}_{max}$} increases, for \mbox{$\text{De} < 1.2$}, the data initially aligns with the Newtonian trend during the increasing portion of the cycle. However, \mbox{$I$} begins to increase at a critical strain rate, leading to the appearance of a clear hysteresis loop in the strain-hardening curves. This transition occurs at lower values of \mbox{$\dot{\varepsilon}_{in}^\prime$} as the maximum Weissenberg number increases.

The 200~ppm and 400~ppm PAA samples display qualitatively similar behavior in both the Lissajous curve of excess pressure drop versus inlet strain rate \mbox{$\dot{\varepsilon}_{out}^\prime(\dot{\varepsilon}_{in}^\prime)$} and the strain-hardening index, as shown in Fig.~S11 of the Supplementary Material.

\begin{figure*}
\centering
\includegraphics[width=\textwidth]{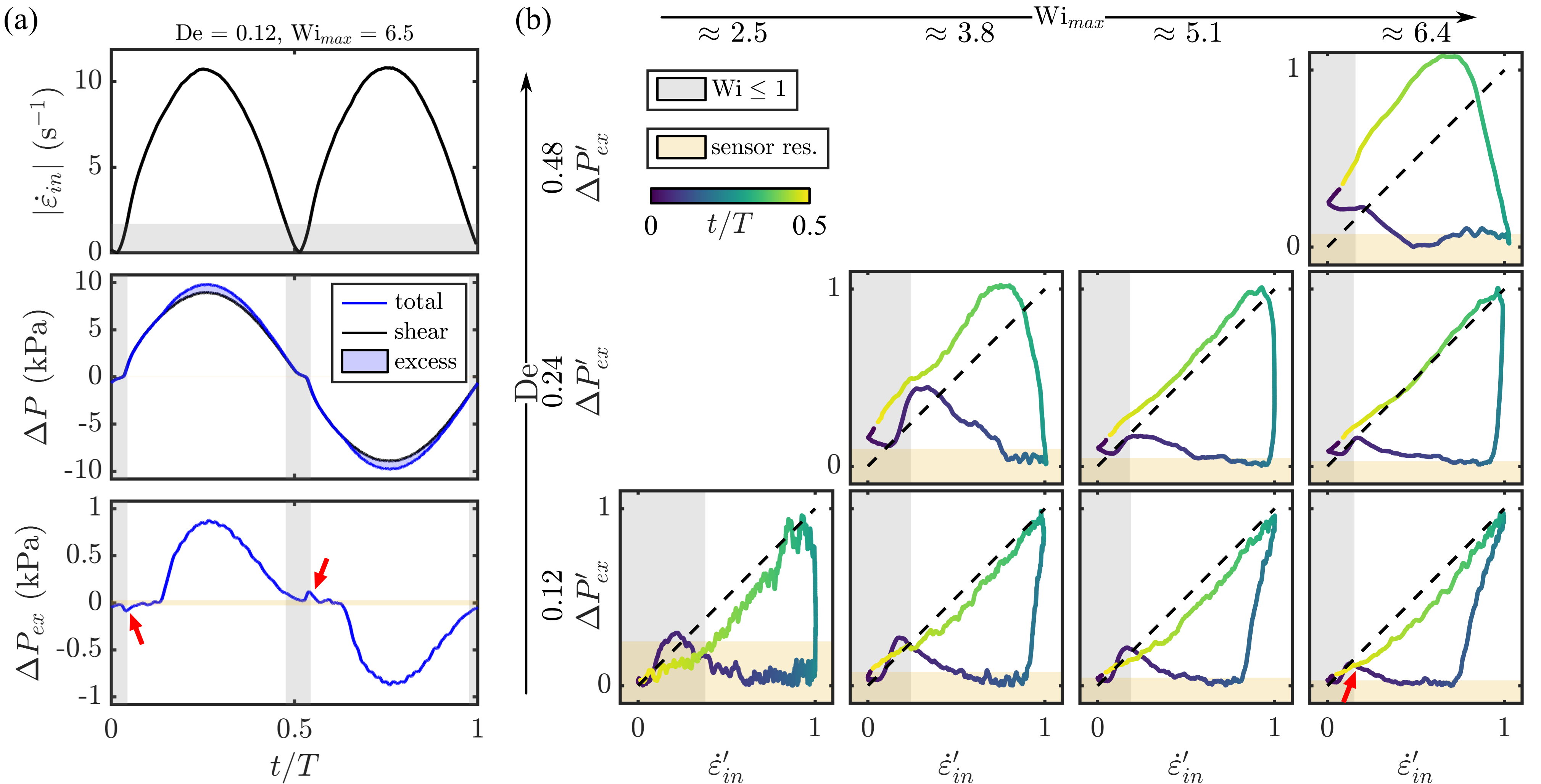}
\caption{Excess pressure during oscillatory LAOE. (a) Representative measurement for the 800~ppm PAA sample showing the absolute inlet extension rate (top), total and shear pressure drops (middle), and excess pressure drop (bottom) over the normalized pulsation cycle. Data in (a) is shown for \mbox{$\text{De}=0.12$} and \mbox{$\text{Wi}_{max}=6.5$} (\mbox{$\dot{\varepsilon}_{0,set}=\unit[10]{s^{-1}}$} and \mbox{$T=\unit[10]{s}$}). (b) Lissajous curves of the normalized excess pressure drop \mbox{$\Delta P_{ex}^\prime$} versus the normalized inlet extension rate \mbox{$\dot{\varepsilon}_{in}^\prime$} for various \mbox{$\text{De}$} and \mbox{$\text{Wi}_{max}$}. Black dashed lines correspond to slopes of 1. Gray areas indicate when \mbox{$\text{Wi}(t) \leq 1$} during the cycle, and yellow areas represent the lower detection limit of the pressure transducer. Red arrows indicate irregularities in the excess pressure drop measurements caused by the reversal of the pump direction. The \mbox{$\text{Wi}_{max}$} values above the panel correspond to approximate values at similar maximum Weissenberg numbers, averaged over the shown \mbox{$\text{De}$} range.}
\label{FIG_OscExcessP}
\end{figure*}

\paragraph{Excess pressure drop}
As with the measurements taken under pulsatile LAOE, we measure the total pressure drop (see Fig.~S12 in the Supplementary Material) and calculate the excess pressure drop \mbox{$\Delta P_{ex}$} to assess the fluid's elastic stress response under oscillatory LAOE. Figure~\ref{FIG_OscExcessP}(a) presents a representative measurement of the total pressure drop, shear pressure drop, and excess pressure drop for the 800~ppm PAA sample at \mbox{$\text{De}=0.12$} and \mbox{$\text{Wi}_{max}=6.5$}. As the imposed strain rate increases, both the total pressure drop and shear pressure drop rise in a similar manner. However, before reaching the maximum strain rate during the cycle, the total pressure drop exceeds the shear pressure drop, resulting in a sudden increase in the excess pressure drop signal at \mbox{$t/T\approx0.125$}. During the descending phase of the sine wave, the excess pressure drop gradually relaxes until it returns to zero. Upon reversing the pump direction, a similar increase and decrease in the excess pressure drop is observed, now in the negative direction. It is important to note that small kinks in the \mbox{$\Delta P_{ex}$} signal appear around \mbox{$\Delta P_{ex}\approx0$} when the pumps change direction, as indicated by the red arrows in Fig.~\ref{FIG_OscExcessP}(a).

In the case of pulsatile LAOE, we observed an increase in the excess pressure drop at the beginning of the cycle (see Fig.~\ref{FIG_PulseExcessP}(a)). In contrast, under oscillatory LAOE, a pronounced plateau where \mbox{$\Delta P_{ex} \approx 0$} appears at the start of the measurement. This difference arises because, under pulsatile flow conditions, the fluid experiences a strain rate of \mbox{$\dot{\varepsilon}_{off}$} at the beginning of the measurement due to the constant background flow, whereas \mbox{$\dot{\varepsilon}_{off} = 0$} during oscillatory LAOE. Consequently, it takes longer for the oscillatory case to reach the critical strain rate necessary to stretch the polymer and contribute to the elastic stress. Once this threshold is surpassed, the excess pressure drop increases and decreases in a similar manner in both LAOE modes.

In Fig.~\ref{FIG_OscExcessP}(b), we show the Lissajous curves of the normalized excess pressure drop \mbox{$\Delta P_{ex}^\prime$} as a function of the normalized inlet extension rate \mbox{$\dot{\varepsilon}_{in}^\prime$} for various \mbox{$\text{De}$} and \mbox{$\text{Wi}_{max}$}. At low \mbox{$\text{De} = 0.12$} and high \mbox{$\text{Wi}_{max} \approx 6.4$}, \mbox{$\Delta P_{ex}^\prime \approx 0$} at the start of the measurement until \mbox{$\dot{\varepsilon}_{in}^\prime \approx 0.7$}, when a sudden increase in the excess pressure drop can be observed until the maximum strain rate is reached. During the descending phase, \mbox{$\Delta P_{ex}^\prime$} follows a slightly concave trend until it reaches zero at \mbox{$t/T = 0.5$}. Decreasing \mbox{$\text{Wi}_{max}$} while keeping \mbox{$\text{De} = 0.12$}, we observe a qualitatively similar trend. However, the increase in \mbox{$\Delta P_{ex}^\prime$} is shifted to larger strain rates at smaller \mbox{$\text{Wi}_{max}$}. Consequently, \mbox{$\Delta P_{ex}^\prime$} rises with a steeper slope toward the maximum as \mbox{$\text{Wi}_{max}$} decreases.

This trend becomes more pronounced when maintaining a fixed maximum Weissenberg number while increasing the Deborah number. At \mbox{$\text{Wi}_{max} \approx 6.4$}, we observe a transition in behavior. Initially, the maximum excess pressure drop occurs before the maximum strain rate during the cycle (\mbox{$\text{De} = 0.12$}). As \mbox{$\text{De}$} increases to \mbox{$0.24$}, the excess pressure drop increases abruptly at the maximum strain rate. Further increasing \mbox{$\text{De}$} to \mbox{$0.48$} leads to a scenario where the maximum excess pressure drop is reached during the descending phase of the sine wave, indicated by a backward-bending slope in the Lissajous curve after reaching the maximum strain rate. This transition resembles the observed change in the Lissajous curve under pulsatile LAOE, which shifts from a concave shape to a linear, and then to a convex shape, surpassing the diagonal line as the excess pressure drop decreases after reaching its maximum (see Fig.~\ref{FIG_PulseExcessP}(b)).

Qualitatively similar Lissajous curves for the excess pressure drop as a function of the inlet strain rate under oscillatory LAOE are also observed for the 200~ppm and 400~ppm PAA samples, as shown in Fig.~S13 in the Supplementary Material.

\subsection{Comparison of LAOE modes}\label{sec_compare}
\begin{figure*}
\centering
\includegraphics[width=\textwidth]{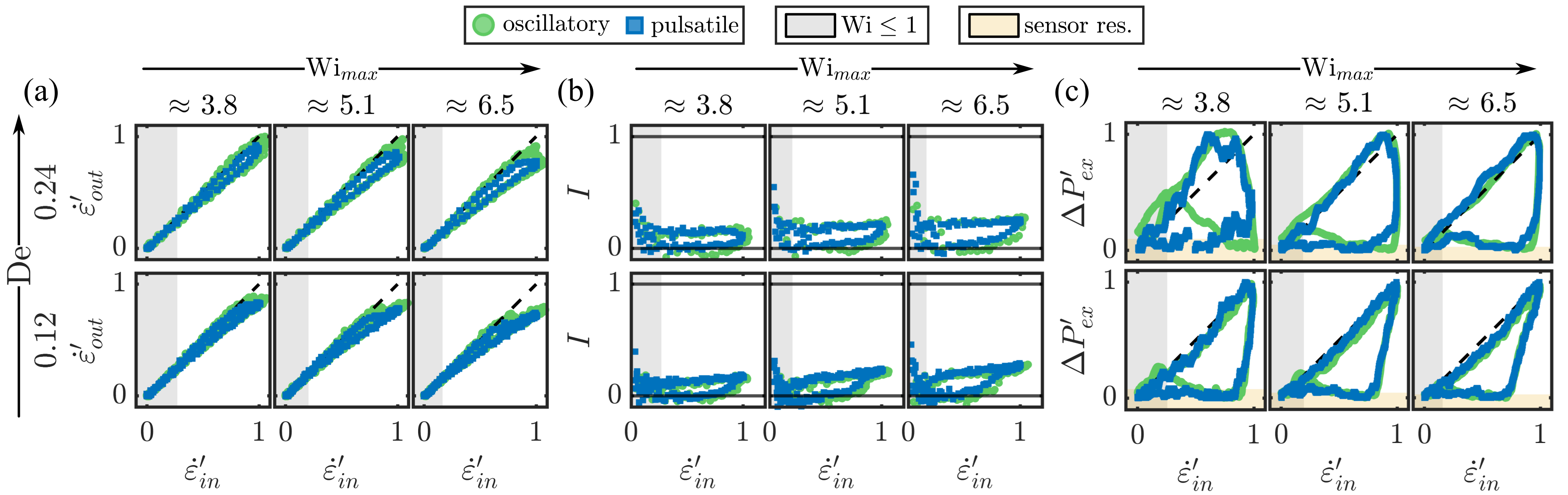}
\caption{Comparison of pulsatile and oscillatory LAOE. (a) Lissajous curves of the normalized strain rate in the outlet direction \mbox{$\dot{\varepsilon}_{out}^\prime$} as a function of the normalized inlet extension rate \mbox{$\dot{\varepsilon}_{in}^\prime$}. (b) Strain-hardening index \mbox{$I=1-\dot{\varepsilon}_{out}^\prime/\dot{\varepsilon}_{in}^\prime$} as a function of the normalized inlet extension rate \mbox{$\dot{\varepsilon}_{in}^\prime$} during the cycle. (c) Lissajous curves of the normalized excess pressure drop \mbox{$\Delta P_{ex}^\prime$} as a function of the normalized inlet extension rate \mbox{$\dot{\varepsilon}_{in}^\prime$}. Data are shown for the 800~ppm PAA solution for various \mbox{$\text{De}$} and \mbox{$\text{Wi}_{max}$}. Black dashed lines represent slopes of 1. Gray areas indicate when \mbox{$\text{Wi}(t)\leq 1$} during the cycle, as calculated from the inlet strain rate. The \mbox{$\text{Wi}_{max}$} values above the panel correspond to approximate values of the maximum Weissenberg number averaged over the shown \mbox{$\text{De}$} range.}
\label{FIG_Combi}
\end{figure*}

We observed distinct deviations from Newtonian flow behavior in the polymeric samples under pulsatile and oscillatory LAOE. To analyze these differences, we compare both LAOE modes using results from PIV and pressure drop measurements obtained during the cycle. Figure~\ref{FIG_Combi} shows (a) the Lissajous curves of the normalized strain rate in the outlet direction \mbox{$\dot{\varepsilon}_{out}^\prime(t)$} as a function of the normalized inlet extension rate \mbox{$\dot{\varepsilon}_{in}^\prime(t)$}, (b) the strain-hardening index \mbox{$I$} as a function of the normalized inlet extension rate \mbox{$\dot{\varepsilon}_{in}^\prime(t)$} and (c) Lissajous curves of the normalized excess pressure drop \mbox{$\Delta P_{ex}^\prime$} over the normalized inlet extension rate \mbox{$\dot{\varepsilon}_{in}^\prime$} for the 800~ppm PAA solution at matching \mbox{$\text{De}$} and \mbox{$\text{Wi}_{max}$}. Note that for oscillatory LAOE, we calculate the Deborah number using half the period duration, as explained in Sec.~\ref{sec_osc}.

Overall, we observe excellent agreement between both LAOE modes. The Lissajous plots of the extension rates and the strain-hardening index show a quantitative overlap at the compared Weissenberg and Deborah numbers, as demonstrated in Fig.~\ref{FIG_Combi}(a) and (b). The deviations from Newtonian flow behavior, specifically, the deviation from the angle bisector in Fig.~\ref{FIG_Combi}(a) and from \mbox{$I=1$} in Fig.~\ref{FIG_Combi}(b), occur at the similar normalized inlet strain rates for both pulsatile and oscillatory LAOE.

In oscillatory LAOE, the strain rate oscillates around zero, whereas a constant background flow is applied in pulsatile LAOE before the experiment begins, as detailed in Sec.~\ref{sec_timeMeasurements}. Additionally, the imposed inlet strain rate follows a sinusoidal signal in pulsatile LAOE, while in oscillatory LAOE, it follows the modulus of a sine wave. Although both signals share the same maximum, minimum, and periodicity, their shapes differ, particularly when the imposed strain rate approaches zero. Despite these differences, the qualitative shapes of the curves are consistent throughout the cycle, suggesting that the critical conditions for the onset of flow modifications in terms of strain rate or Weissenberg number are independent of the LAOE mode. These differences in imposed strain rates primarily occur at small strain rates (\textit{i.e.}, small temporal Weissenberg numbers). During this part of the cycle, no pronounced deviation from the Newtonian case is observed for either LAOE mode. Therefore, the exact strain rate profile (sinusoidal or modulus of sine) does not significantly affect the results shown in Fig.~\ref{FIG_Combi}.

Furthermore, the Lissajous curves of the excess pressure drop and inlet strain rate exhibit good agreement, displaying similar shapes and evolution with \mbox{$\text{De}$} and \mbox{$\text{Wi}_{max}$}, as shown in Fig.~\ref{FIG_Combi}(c). Due to the reversal of the flow direction and the change in pump operation mode from pumping to withdrawing, small deviations from the desired sinusoidal signal are observed when the strain rate passes through zero in the oscillatory measurements. These irregularities are absent in the pulsatile measurements, which is an advantage of pulsatile LAOE over oscillatory LAOE.

Based on the comparison above, we conclude that both LAOE modes provide similar results in determining the nonlinear stress response of the investigated polymer solutions. However, they differ in terms of experimental implementation and operational constraints. From a technical perspective, the pulsatile approach is easier to realize experimentally. One of the primary advantages of pulsatile LAOE is that the pumps do not need to reverse direction, which simplifies the experimental setup and reduces the complexity associated with the pump control. Another significant advantage of pulsatile LAOE is the ability to avoid the problem of reaching the minimum flow rate that the pumps can deliver. By choosing a constant background strain rate slightly larger than the set strain rate amplitude, the system ensures that the pumps do not approach their minimum flow rate limit during the cycle. This strategy effectively prevents the plateau that can occur in the inlet strain rate and pressure drop signals, which can be problematic at low strain rate amplitudes (see Fig.~\ref{FIG_OscNewton}(e)). Moreover, pulsatile LAOE can be optimized by combining it with an adapted input signal that accounts for the specific characteristics of the system~\cite{Recktenwald2021b}. This adaptation would allow us to impose a perfect sinusoidal strain rate signal, even at higher pulsation frequencies, without suffering from distortions caused by pump reversal or flow rate limitations.

\subsection{Onset of non-linearities in LAOE}\label{sec_nlOnset}
To better understand the nonlinear viscoelastic response, we determine the critical conditions for the onset of nonlinear behavior of the flow field during LAOE. As shown in Fig.\ref{FIG_PulseNonNewton}(d) and Fig.\ref{FIG_OscNonNewton}(c), the outlet strain rate \mbox{$\dot{\varepsilon}_{out}$} begins to fall below the inlet strain rate \mbox{$\dot{\varepsilon}_{in}$} as the maximum Weissenberg number increases. This raises the question: what critical interplay between flow strength \mbox{$\text{Wi}_{max}$} and cycle frequency \mbox{$\text{De}$} drives this transition under LAOE?

\begin{figure}
\centering
\includegraphics[width=.5\textwidth]{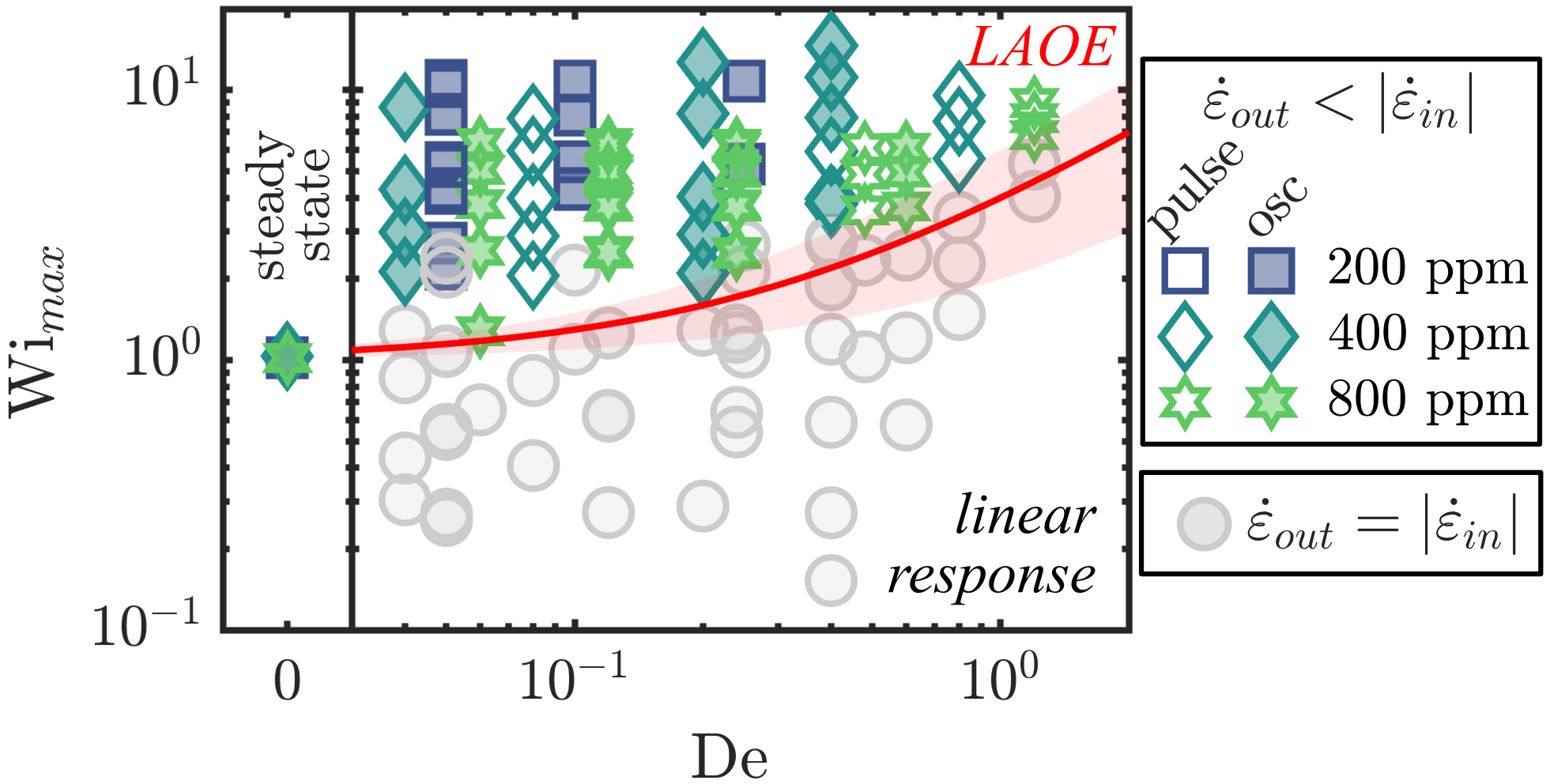}
\caption{Critical conditions for the onset of non-linearities during LAOE. Pipkin space representing all experiments at different \mbox{$\text{De}$} and \mbox{$\text{Wi}_{max}$} combinations. Colored symbols indicate measurements where the outlet extension rate drops below the inlet rate during the cycle. Open symbols show measurements under pulsatile LAOE and filled symbols under oscillatory LOAE. Gray circles represent combinations of \mbox{$\text{De}$} and \mbox{$\text{Wi}_{max}$} that did not show a deviation of \mbox{$\dot{\varepsilon}_{out}$} from \mbox{$\dot{\varepsilon}_{in}$} during the cycle. The data points at \mbox{$\text{De}=0$} represent the steady-state data. The red line corresponds to the equation of \mbox{$\text{Wi}_{c}^{LAOE}=\text{Wi}_{c}(k\text{De}+1)$} with \mbox{$k=3$}, as described in the main text. The shaded red area indicated solutions for \mbox{$1\leq k\leq5$}.}
\label{FIG_PD}
\end{figure}

Figure~\ref{FIG_PD} presents a Pipkin space representation of the experimental results across varying \mbox{$\text{De}$} and \mbox{$\text{Wi}_{max}$} combinations. Gray circles represent conditions where \mbox{$\dot{\varepsilon}_{out}=\lvert\dot{\varepsilon}_{in}\rvert$}, while colored symbols represent cases where the outlet extension rate dropped below the inlet rate during the cycle. Open symbols correspond to pulsatile LAOE, and filled symbols indicate oscillatory LAOE measurements. Note that these results are based solely on the PIV measurements.

To quantify this transition, we introduce an effective Weissenberg number \mbox{$\text{Wi}_{eff}$} for periodic extensional flow

\begin{equation}
    \text{Wi}_{eff}=\frac{\text{Wi}_{max}}{k\text{De}+1},
    \label{eq_Wieff} 
\end{equation}

\noindent which is inspired by Zhou~and~Schroeder~\cite{Zhou2016a}, who employed a similar definition to describe the transition from linear to nonlinear responses in single DNA molecules under oscillatory extensional flow. The constant \mbox{$k$} acts as a prefactor, previously used to align data from Hookean dumbbell models and numerical simulations. In their study, Zhou~and~Schroeder~\cite{Zhou2016a} identified a critical flow strength for the transition from linear to nonlinear responses at \mbox{$\text{Wi}_{eff} = 0.5$}.

In the experiments presented in this study, we observe that in the limit of steady flow conditions (\mbox{$f \rightarrow 0$}), the pulsation frequency approaches zero, causing the Deborah number \mbox{$\text{De}$} to also tend to zero. Consequently, the effective Weissenberg number reduces to \mbox{$\text{Wi}_{eff} = \text{Wi}_{max}$}, which corresponds to the Weissenberg number \mbox{$\text{Wi}$} under steady conditions. As shown in Fig.~\ref{FIG_SteadyFlow}(g), the outlet strain rate \mbox{$\dot{\varepsilon}_{out}$} falls below the inlet strain rate \mbox{$\dot{\varepsilon}_{in}$} under steady flow at a critical Weissenberg number \mbox{$\text{Wi}_{c} \approx 1$} (\mbox{$\text{De} = 0$} in Fig.~\ref{FIG_PD}). Hence, under steady extension, \mbox{$\text{Wi}_{eff} = \text{Wi}_{c} = 1$}.

Rearranging Eq.~\ref{eq_Wieff} gives \mbox{$\text{Wi}_{max}(\text{De})=\text{Wi}_{eff}(k\text{De}+1)$}. Relating the steady flow conditions with \mbox{$\text{Wi}_{eff} = \text{Wi}_{c}$} leads to \mbox{$\text{Wi}_{c}^{LAOE}=\text{Wi}_{c}(k\text{De}+1)$}, where \mbox{$\text{Wi}_{c}^{LAOE}$} is the critical flow strength during LAOE. The red line in Fig.~\ref{FIG_PD} represents this relation, using a prefactor of \mbox{$k = 3$}, which provides a reasonable fit to the experimental data. The shaded red area shows the variation in the critical onset for \mbox{$1 \leq k \leq 5$}, indicating the sensitivity of the results to the prefactor \mbox{$k$} especially at high \mbox{$\text{De} > 1$}.

This critical relation \mbox{$\text{Wi}_{c}^{LAOE}(\text{De})$} separates the experimental results in Newtonian-like linear response and nonlinear behavior under LAOE. In the linear regime, polymers are only slightly perturbed beyond equilibrium. By contrast, increasing the flow strength into the nonlinear regime results in a large deformation and higher degree of stretching, depending on \mbox{$\text{De}$}. For \mbox{$\text{De} < 1$}, the polymer molecules are stretched since the cycle period \mbox{$T$} exceeds the polymer relaxation time \mbox{$\lambda$}. When the pulsation frequency increases beyond \mbox{$\text{De} > 1$}, Zhou~\textit{et al.}~\cite{Zhou2016a} described nonlinear and linear unsteady regimes for DNA stretching dynamics under highly unsteady flow conditions. However, due to technical constraints of the pumps used in this study, our experiments are restricted to \mbox{$\text{De} \lesssim 1$}. Note that the findings in Fig.~\ref{FIG_PD} are also in agreement with the results shown in Fig.~\ref{FIG_PulseNonNewton}(c), where a slight shift in the ratio of \mbox{$\lvert\dot{\varepsilon}_{out}/\dot{\varepsilon}_{in}\rvert$} towards higher strain rate amplitudes (higher \mbox{$\text{Wi}_{max}$}) with increasing \mbox{$\text{De}$} is observed.

Based on the critical extension rate for the onset of nonlinear behavior (\mbox{$\dot{\varepsilon}_{out}<\lvert\dot{\varepsilon}_{in}\rvert$}, shown in Fig.\ref{FIG_PulseNonNewton}(d) and Fig.\ref{FIG_OscNonNewton}(c)), we calculate the accumulated fluid strain \mbox{$\varepsilon_{c}$}. The strain required to transition from linear to nonlinear behavior is found to be in the range of \mbox{$\varepsilon_{c}=2-6$}. This range aligns well with previously reported results on the uncoiling of polymers before reaching a steady-state extension during filament stretching rheometry~\cite{McKinley2002a}.

\subsection{Comparison between experiment and simulation}\label{sec_compareSim}
\begin{figure*}
\centering
\includegraphics[width=\textwidth]{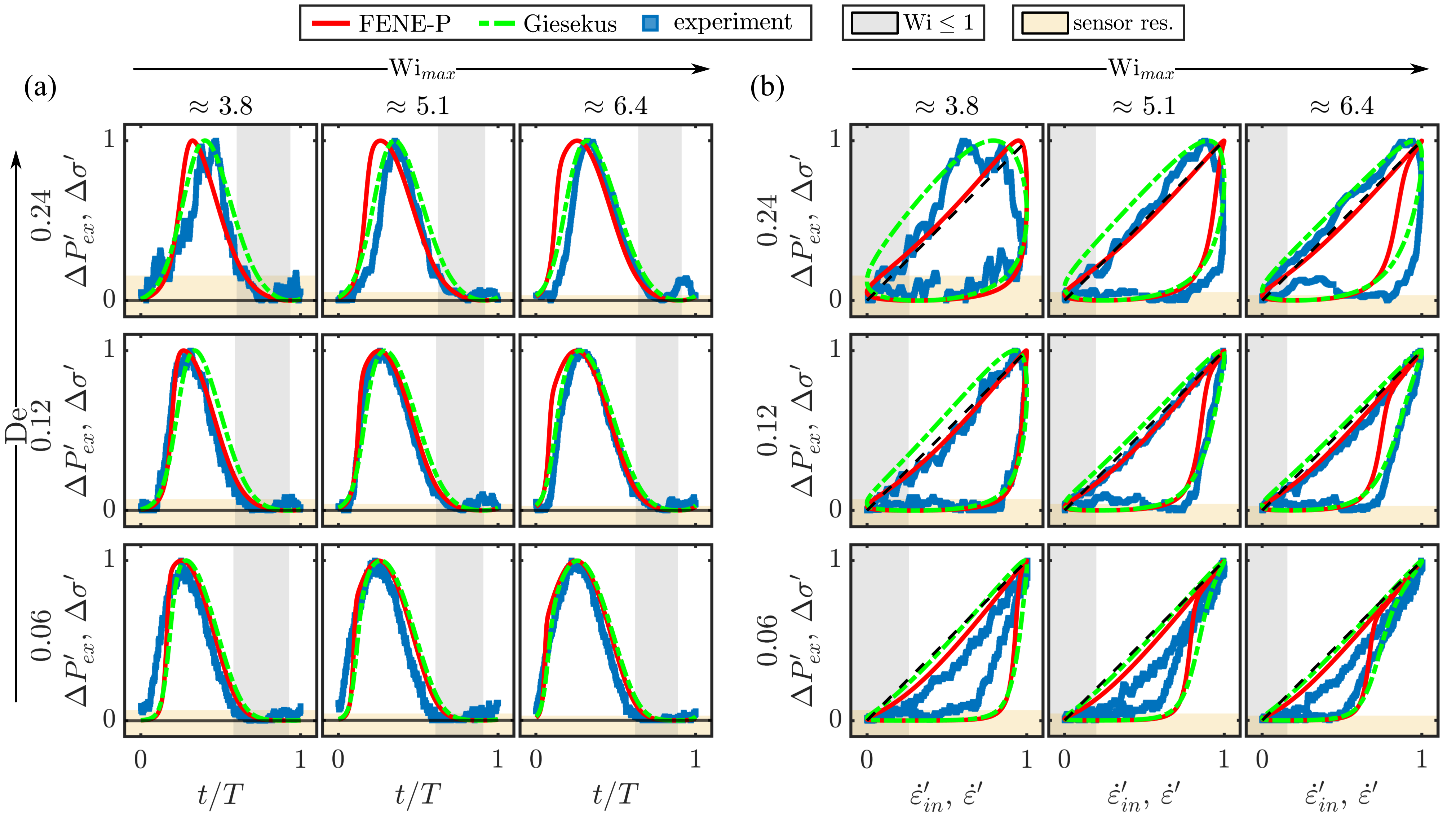}
\caption{Comparison of simulation and experiment for pulsatile LAOE. (a) Temporal evolution of the normalized stress \mbox{$\Delta \sigma^\prime$} in simulations and the normalized excess pressure drop \mbox{$\Delta P_{ex}^\prime$} in experiments during the normalized cycle. (b) Lissajous curves of the normalized excess pressure drop \mbox{$\Delta P_{ex}^\prime$} as a function of the normalized inlet extension rate \mbox{$\dot{\varepsilon}_{in}^\prime$} for the experiment, and Lissajous curves of the normalized stress \mbox{$\Delta \sigma^\prime$} as a function of the imposed normalized extension rate \mbox{$\dot{\varepsilon}^\prime$} for the simulations. Experimental data are shown for the 800~ppm PAA solution at various \mbox{$\text{De}$} and \mbox{$\text{Wi}_{max}$}. Numerical simulations are performed using the FENE-P and Giesekus models at matching \mbox{$\text{De}$} and \mbox{$\text{Wi}_{max}$} values.}
\label{FIG_CombiSim}
\end{figure*}

In addition to the experimental results from the LAOE implementation in the OSCER, we perform simulations to further validate the challenging experimental setup. We compare the results from pulsatile LAOE with the responses of two viscoelastic models under a homogeneous planar extensional flow using two different models, FENE-P and Giesekus. Simulations are conducted by matching Deborah and Weissenberg numbers, using the rheological parameters of the 800~ppm PAA solution. We calculate the first normal stress difference \mbox{$N_1$}, as detailed in Sec.~\ref{sec_methodsim}, and compare it with the excess pressure drop measured in the experiments. The maximum Weissenberg numbers, corresponding to the maximum strain rate along the inlet axis, are used as input for the simulations. Figure~\ref{FIG_CombiSim} compares the normalized excess pressure drop from the experiments with the normalized stress difference \mbox{$\Delta \sigma = N_1 / N_{1,max}$} from the simulations during (a) the normalized period and (b) as a function of the inlet strain rate for various \mbox{$\text{De}$} and \mbox{$\text{Wi}_{max}$}.

Both the FENE-P and Giesekus models exhibit a qualitatively similar stress response, with the stress increase and decrease occurring slightly faster in the FENE-P model. Good qualitative agreement is observed between the temporal stress evolution in the experimental data and both simulations at comparable Deborah and Weissenberg numbers. The maximum stress is reached at approximately the same point in the cycle in both simulations and experiments. For example, at \mbox{$\text{Wi}_{max} \approx 6.4$} and \mbox{$\text{De} = 0.12$} in Fig.~\ref{FIG_CombiSim}(a), the Giesekus model closely matches the experimental results. Additionally, the Lissajous curves of normalized stress as a function of strain rate show triangular shapes similar to those observed in the experimental data, as shown in Fig.~\ref{FIG_CombiSim}(b). However, due to noise in the experimental data, it is not possible to definitively determine which model provides a better fit to the experimental data.

In the experiments, we observed a transition in the shape of the Lissajous curves during the receding phase of the sine wave, following the maximum stress, as the Deborah number (\mbox{$\text{De}$}) increased (see Fig.\ref{FIG_PulseExcessP}(b)). Specifically, the shape evolves from concave to linear and then to convex. Notably, this transition is only partially captured by the two models. At low \mbox{$\text{De}$}, the FENE-P model predicts a slightly more pronounced concave shape during the stress decrease compared to the Giesekus model. However, as \mbox{$\text{De}$} increases, the Giesekus model predicts the full transition from concave to linear to convex shapes in agreement with our experiments. In contrast, the FENE-P model does not exhibit convex shapes during the stress reduction phase of the cycle (see black dashed line in Fig.\ref{FIG_CombiSim}(b)).

Taken together, the numerical simulations exhibit a similar temporal stress buildup and decay during the pulsation cycle, as well as comparable Lissajous curves of normalized stress versus normalized extension rate, consistent with the experimental results at comparable Deborah and Weissenberg numbers. However, we note that the current analytical model assumes a homogeneous extensional flow and does not account for flow modifications induced by polymer stretching and strain hardening in the OSCER. Future spatially resolved simulations (\textit{i.e.}, CFD) could address whether such phenomena occur during LAOE in numerical simulations and evaluate their potential impact on the results presented in Fig.~\ref{FIG_CombiSim}.


\section{Conclusions}
This study introduces an experimental approach to investigate the response of complex fluids of low viscosity under LAOE. We employed a microfluidic OSCER device to generate a homogeneous planar extensional flow. The time-dependent flow field within the OSCER was analyzed using micro-particle image velocimetry, while the simultaneous excess pressure drop \mbox{$\Delta P_{ex}$} was measured to evaluate the fluid's elastic stress response.  

We examined the time-dependent flow of a Newtonian fluid during LAOE, exploring a range of oscillation amplitudes and frequencies, and demonstrated the linearity between the applied strain rate and the measured pressure drop both under pulsatile and oscillatory LAOE. 

In contrast, we observed strong deviations from the Newtonian flow behavior for dilute polymer solutions under LAOE. During the LAOE cycle, the average strain rate along the extension axis \mbox{$\dot{\varepsilon}_{out}$} falls below the average strain rate along the compression axis \mbox{$\dot{\varepsilon}_{in}$} above a critical strain rate. The magnitude of this phenomenon increases with increasing maximum Weissenberg number or flow strength. Concurrently, a significant excess pressure drop arises, and we reveal non-trivial Lissajous curves of the excess pressure drop \mbox{$\Delta P_{ex}$} as a function of the imposed strain rate \mbox{$\dot{\varepsilon}_{in}$}. At comparable Deborah and Weissenberg numbers, the oscillating and pulsatile sinusoidal LAOE modes result in quantitatively similar results regarding time-dependent flow field modifications and elastic stress response. We identify the critical conditions in terms of \mbox{$\text{Wi}_{c}^{LAOE}(\text{De})$} for the onset of non-linearities under oscillatory extension, which seem to be identical under pulsatile and oscillatory LAOE. 

Ultimately, we validate our experimental findings with preliminary numerical predictions assuming a homogeneous flow. Good qualitative agreement is observed between the temporal stress evolution in the experimental data and both simulations at comparable Deborah and Weissenberg numbers. 

Focusing on the experimental realization of LAOE for dilute polymer solutions and low-viscosity fluids, we introduced a comprehensive framework to measure the dynamic nonlinear stress response during planar extension. In this study, we concluded with measurements of the excess pressure drop \mbox{$\Delta P_{ex}$} during the cycle.

Future work is needed to establish a robust and physically meaningful framework for understanding the nonlinear stress response during LAOE, similar to the various analytical methods developed for LAOS~\cite{Wilhelm1998, Ewoldt2008, Rogers2012, Lee2019a}. Further analysis of hysteresis loops and deviations from sinusoidal behavior, including higher-order harmonics, could provide deeper insights into the underlying process during dynamic extension. Translating Lissajous curves of excess pressure drop \mbox{$\Delta P_{ex}$} versus strain rate into excess pressure drop as a function of strain may help quantify energy dissipation during LAOE, analogous to LAOS~\cite{Ewoldt2010}. Moreover, combining LAOS and LAOE data will enable probing nonlinear dynamics, support material design applications, and assist in selecting and validating constitutive models for complex materials.

In this work, we measured the excess pressure drop to quantify the fluid's elastic stress. However, our setup can be adapted to study nonlinear stress dynamics in other complex fluids that do not generate large excess pressures. Combining the proposed LAOE method with time-dependent birefringence measurements could further enable the quantification of molecular or particle orientation in polymeric~\cite{Haward2010b, Haward2011a} or rod-like systems~\cite{Calabrese2023, Calabrese2024b} exhibiting flow-induced birefringence during extension.

In summary, our study introduces a promising new methodology for characterizing complex fluids under controlled nonlinear transient flow conditions, providing valuable insights into their behavior in dynamic and extensional regimes.

\section*{Acknowledgements}
S.M.R., A.Q.S., and S.J.H. gratefully acknowledge the support of the Okinawa Institute of Science and Technology Graduate University (OIST) with subsidy funding from the Cabinet Office, Government of Japan, and also funding from the Japan Society for the Promotion of Science (JSPS, Grant No. 24K17736, 24K00810, and 24K07332).

\section*{Declaration of interests}
The authors declare that they have no known competing financial interests or personal relationships that could have appeared to influence the work reported in this paper.

\section*{Data Availability}
All study data are included in the article and/or supporting information.

\appendix
\section{Full system of ODEs}\label{fullodes}

The full system of ODEs for the FENE-P model under 2D pulsatile flow in a normalized form is given by 

\begin{subequations}
\begin{align}
\displaystyle{\frac{\dif A_{xx}}{\dif t}} = &\displaystyle{\frac{1}{2}}\displaystyle{\frac{\text{Wi}_\text{max}}{\text{De}}}A_{xx} (1+\sin(2\pi t)) \\& \nonumber -  \displaystyle{\frac{1}{\text{De}}} \left(\displaystyle{\frac{L^2}{L^2-A_{xx}-A_{yy}-A_{zz}}}A_{xx} - \displaystyle{\frac{L^2}{L^2-3}}\right),
\end{align}
\begin{align}
\displaystyle{\frac{\dif A_{yy}}{\dif t}} = &-\displaystyle{\frac{1}{2}}\displaystyle{\frac{\text{Wi}_\text{max}}{\text{De}}}A_{yy} (1+\sin(2\pi t)) \\& \nonumber - \displaystyle{\frac{1}{\text{De}}} \left(\displaystyle{\frac{L^2}{L^2-A_{xx}-A_{yy}-A_{zz}}}A_{yy} - \displaystyle{\frac{L^2}{L^2-3}}\right),
\end{align} 
\begin{align}
\displaystyle{\frac{\dif A_{zz}}{\dif t}} = - \displaystyle{\frac{1}{\text{De}}} \left(\displaystyle{\frac{L^2}{L^2-A_{xx}-A_{yy}-A_{zz}}}A_{zz} - \displaystyle{\frac{L^2}{L^2-3}}\right),
\end{align} \label{eq:fenepFull}
\end{subequations}

\noindent and those for the Giesekus model are given by

\begin{subequations}
\begin{align}
\displaystyle{\frac{\dif A_{xx}}{\dif t}} = &  \displaystyle{\frac{1}{2}}\displaystyle{\frac{\text{Wi}_\text{max}}{\text{De}}}A_{xx} (1+\sin(2\pi t)) \\ & \nonumber - \displaystyle{\frac{1}{\text{De}}}\left[1+\alpha(A_{xx}-1)\right](A_{xx}-1),
\end{align} 
\begin{align}
\displaystyle{\frac{\dif A_{yy}}{\dif t}} = & -\displaystyle{\frac{1}{2}}\displaystyle{\frac{\text{Wi}_\text{max}}{\text{De}}}A_{yy} (1+\sin(2\pi t)) \\ & \nonumber - \displaystyle{\frac{1}{\text{De}}}\left[1+\alpha(A_{yy}-1)\right](A_{yy}-1),
\end{align} 
\begin{align}
\displaystyle{\frac{\dif A_{zz}}{\dif t}} = - \displaystyle{\frac{1}{\text{De}}}\left[1+\alpha(A_{zz}-1)\right](A_{zz}-1),
\end{align}  \label{eq:giesekusFull}
\end{subequations}

\noindent and the components of the extra stress tensor are recovered by 

\begin{subequations}
\begin{align}
\tau_{xx} = & \frac{(1-\beta)}{\text{Wi}_\text{max}}\left(\frac{L^2}{L^2-A_{xx}-A_{yy}-A_{zz}}A_{xx} - \frac{L^2}{L^2-3} \right) \\& \nonumber + \frac{1}{2}\beta(1+\sin(2\pi t)),
\end{align}
\begin{align}
\tau_{yy} = & -\frac{(1-\beta)}{\text{Wi}_\text{max}}\left(\frac{L^2}{L^2-A_{xx}-A_{yy}-A_{zz}}A_{yy} - \frac{L^2}{L^2-3} \right) \\& \nonumber - \frac{1}{2}\beta(1+\sin(2\pi t)),
\end{align}
\begin{align}
\tau_{zz} = \frac{(1-\beta)}{\text{Wi}_\text{max}}\left(\frac{L^2}{L^2-A_{xx}-A_{yy}-A_{zz}}A_{zz} - \frac{L^2}{L^2-3} \right),
\end{align}\label{eq:fenepStressFull}
\end{subequations}

for the FENE-P model, and by

\begin{subequations}
\begin{equation}
\tau_{xx} = \frac{(1-\beta)}{\text{Wi}_\text{max}}\left(A_{xx} - 1 \right) + \frac{1}{2}\beta(1+\sin(2\pi t)),
\end{equation}
\begin{equation}
\tau_{yy} = \frac{(1-\beta)}{\text{Wi}_\text{max}}\left(A_{yy} - 1 \right) - \frac{1}{2}\beta(1+\sin(2\pi t)),
\end{equation}
\begin{equation}
\tau_{zz} = \frac{(1-\beta)}{\text{Wi}_\text{max}}\left(A_{zz} - 1 \right),
\end{equation} \label{eq:GiesekusStressFull}
\end{subequations}

\noindent for the Giesekus model. We note that the ODEs for the Giesekus model are uncoupled, whereas those for the FENE-P model are coupled through the extensibility function. This coupling in the FENE-P model is weak however due to the dominance of one normal component of $\tens{A}$ when the strain rate is high.  Also note that with the initial condition that $\tens{A} = \tens{I}$ (\textit{i.e.}, initially at equilibrium), it is the case that $A_{zz} = 1$ for all times for the Giesekus model, but $A_{zz} \neq 1$ for all times in the FENE-P model. This does not have a significant impact on the response of the FENE-P model however.

\bibliographystyle{elsarticle-num} 
\bibliography{main}

\end{document}